\documentclass[12pt]{iopart}

\usepackage{iopams}
\usepackage{verbatim} %mehrzeilige Kommentare
\usepackage[utf8]{inputenc}

\usepackage{graphicx}
\usepackage[colorlinks=true,citecolor=blue]{hyperref}
\usepackage{units}

\newcommand{\ket}[1]{|#1\rangle}
\renewcommand{\vec}[1]{\mathbf{#1}}
\newcommand{\text}[1]{\mathsf{#1}}

\def\up{\uparrow}
\def\down{\downarrow}
\def\kB {k_\text{B}}
\newcommand{\kBT}{k_\mathsf{B}T}

\begin{document}
\topical{Thermoelectric energy harvesting with quantum dots}

\author{Björn Sothmann$^1$, Rafael Sánchez$^2$ and Andrew N. Jordan$^3,4$}

\address{$^1$D\'epartement de Physique Th\'eorique, Universit\'e de Gen\`eve, CH-1211 Gen\`eve 4, Switzerland}
\address{$^2$Instituto de Ciencia de Materiales de Madrid (ICMM-CSIC), Cantoblanco, E-28049 Madrid, Spain}
\address{$^3$Department of Physics and Astronomy, University of Rochester, Rochester, New York 14627, USA}
\address{$^4$Institute for Quantum Studies, Chapman University - Orange, CA 92866, USA}

\ead{bjorn.sothmann@unige.ch}
\begin{abstract}
We review recent theoretical work on thermoelectric energy harvesting in multi-terminal quantum-dot setups. We first discuss several examples of nanoscale heat engines based on Coulomb-coupled conductors. In particular, we focus on quantum dots in the Coulomb-blockade regime, chaotic cavities and resonant tunneling through quantum dots and wells. We then turn towards quantum-dot heat engines that are driven by bosonic degrees of freedom such as phonons, magnons and microwave photons. These systems provide interesting connections to spin caloritronics and circuit quantum electrodynamics.
\end{abstract}

%Uncomment for PACS numbers title message
% 73.63.-b Electronic transport in nanoscale materials and structures
% 73.23.-b Electronic transport in mesoscopic systems
% 73.40.-c Electronic transport in interface structures
% 73.50.Lw Thermoelectric effects
% 85.80.Fi Thermoelectric devices
% 05.60.-k Transport processes
% 07.20.Pe Heat engines; heat pumps; heat pipes

% Keywords required only for MST, PB, PMB, PM, JOA, JOB? 
%\vspace{2pc}
%\noindent{\it Keywords}: Article preparation, IOP journals
% Uncomment for Submitted to journal title message
\submitto{\NT}
% Comment out if separate title page not required
\maketitle

\section{\label{sec:intro}Introduction}
Thermoelectric effects have generated an immense interest for quite some time already because they offer the possibility to convert heat from the environment into electrical work~\cite{shakouri_recent_2011,radousky_energy_2012}. This form of energy harvesting is potentially useful for electric circuits on modern computer chips that produce large amounts of heat and currently need to be cooled actively in order not to overheat. One can also use the harvested energy to run auxiliary circuits such as autonomous sensors or recycle the lost energy to charge a battery. Unfortunately, even after decades of material research current thermoelectric materials still have a very low efficiency in converting heat into electrical work and deliver only moderate powers. For this reason, thermoelectric energy harvesting so far is restricted to certain niche applications such as in interplanetary spaceships where the fact that a thermoelectric generator does not require any moving parts and therefore no maintenance turns out to be useful.

We give a brief summary of how conventional thermoelectric devices work to orient the reader. The building blocks of thermoelectricity are the Peltier and Seebeck effect.  The Seebeck effect is the flow of electrical current in response to an applied temperature difference, while the Peltier effect is the reverse: the creation of temperature difference in response to an applied electrical voltage. This is intuitively understood by the fact that the hotter charge carriers diffuse faster than the cold ones, creating a flow of thermal energy from hot to cold, and consequently the imbalanced charge builds up an electrical voltage across the material.

One challenge in making good thermoelectric devices is that materials that are good electrical conductors tend also to be good thermal conductors.  This fact makes it difficult to maintain the necessary temperature difference needed to produce electrical power from the Seebeck effect.  The need for ways to create systems with high electrical conductance, while maintaining low thermal conductance is an outstanding challenge in this field that the interface-based devices we shall describe can help to solve.

Commercial devices use doped semiconductors such as bismuth telluride. In an n-type semiconductor, heat flow is in the same direction of the electron (current) flow, however, in a p-type semiconductor, heat and current flows are in opposite directions. This is useful because a thermoelectric architecture can be built of alternating p-type and n-type semiconductor elements that are connected electrically in series (via metallic contacts) and thermally in parallel. This permits the small electrical voltage (or power) produced by a single element to be increased by the number of elements in the device, making a practically useful system. These are placed between ceramics plates which conduct heat, but not electricity, so one side can be heated and the other cooled to produce the power. Alternatively, power can be applied to cool the cold side, and heat the hot one, acting as a refrigerator.

Mesoscopic solid-state physics can help to overcome the limitations of current thermoelectric materials by providing powerful and highly efficient heat engines operating at the nanoscale. Soon after initial studies on the thermopower of basic mesoscopic structures like quantum point contacts~\cite{streda_quantised_1989,molenkamp_quantum_1990} and quantum dots~\cite{beenakker_theory_1992,staring_coulomb-blockade_1993,humphrey_reversible_2002}, there were first proposals pointing out that structures of reduced dimension can give rise to an increased thermoelectric figure of merit $ZT$ as compared to bulk structures made from the same materials~\cite{hicks_effect_1993,hicks_thermoelectric_1993}. Similar ideas where brought forward by Mahan and Sofo~\cite{mahan_best_1996} who showed that sharp spectral features give rise to high thermoelectric performance as characterized by a high value of $ZT$. Nanoscale conductors such as quantum dots naturally provide these sharp spectral features. Hence, they are promising candidates for thermoelectric energy harvesters.

Initial experiments on quantum dots defined in a two-dimensional electron gas showed saw-toothlike oscillations of the thermopower as a function of gate voltage in agreement with theory~\cite{staring_coulomb-blockade_1993,dzurak_thermoelectric_1997}. Further studies have been carried out on open quantum dots~\cite{godijn_thermopower_1999} and carbon nanotubes~\cite{small_modulation_2003,llaguno_observation_2003}. More recently, the thermopower of quantum dots defined in nanowires has been studied~\cite{svensson_lineshape_2012,svensson_nonlinear_2013}. Interestingly, these experiments could observe nonlinear thermoelectric effects~\cite{svensson_nonlinear_2013}. In another series of experiments the thermopower due to sequential tunneling, cotunneling and the Kondo effect has been investigated in gate-defined quantum dots~\cite{scheibner_thermopower_2005,scheibner_sequential_2007}. The thermopower of double dots was observed in Ref.~\cite{thierschmann_diffusion_2013}.

Two-terminal geometries using mesoscopic conductors have been considered, notably using quantum dots~\cite{beenakker_theory_1992,andreev_coulomb_2001,boese_thermoelectric_2001,humphrey_reversible_2002,turek_cotunneling_2002,koch_thermopower_2004,kubala_quantum-fluctuation_2006,kubala_violation_2008,dubi_thermospin_2009,esposito_thermoelectric_2009,swirkowicz_thermoelectric_2009,billings_signatures_2010,leijnse_nonlinear_2010,costi_thermoelectric_2010,nakpathomkun_thermoelectric_2010,karlstrom_increasing_2011,mani_nanoscale_2011,liu_role_2011,rejec_spin_2012,harbola_quantum_2012,muralidharan_performance_2012,dutt_strongly_2013,muralidharan_thermoelectric_2013,weymann_spin_2013,kennes_interacting_2013,kuo_thermoelectric_2013}. In the two-terminal geometry, both temperature and voltage bias are applied to the sample and the thermoelectric response is investigated. It has the advantage of being the analogue of the traditional thermocouple which has wide applications, mainly in its role as a thermometer. However, for the purpose of energy harvesting, it suffers from the fact that different parts of the same electrical circuit must be at different temperatures which makes thermal isolation difficult.

\begin{figure}
	\centering\includegraphics[width=.45\columnwidth]{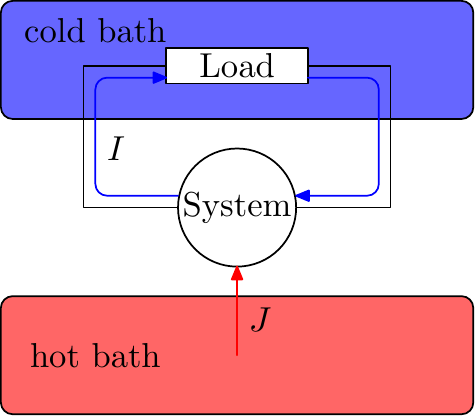}
	\caption{\label{fig:harvest}Sketch of a generic thermoelectric energy harvester: a mesoscopic system situated at the interface between a cold and a hot bath converts a heat current $J$ into a charge current $I$ that is able to power a load or charge a battery.}
\end{figure}
% \textcolor{red}{$\rightarrow$}
In contrast to the above mentioned works based on two-terminal configurations, the problem of a device that powers a circuit by harvesting energy from the outside environment demands a three-terminal geometry, cf. figure~\ref{fig:harvest}. Two terminals at the same temperature define the conductor that supports a charge current $I$. The third terminal represents the coupling to the heat source from which energy but not charge is absorbed. The harvester converts the heat current $J$ into useful power, $P=IV$, that runs a load resistance, represented here by a voltage drop $V$ opposed to the generated current. The properties of the mesoscopic region to which the three terminals are connected define the characteristics of the heat engine, in particular its efficiency, $\eta=P/J$.
% \textcolor{red}{$\leftarrow$}

Only recently has the investigation of thermoelectric effects in quantum-dot structures in three-terminal geometries generated a lot of attention~\cite{sanchez_optimal_2011,sothmann_rectification_2012,jordan_powerful_2013,sanchez_detection_2012,sothmann_powerful_2013,sanchez_correlations_2013,rutten_reaching_2009,entin-wohlman_three-terminal_2010,entin-wohlman_three-terminal_2012,jiang_thermoelectric_2012,jiang_three-terminal_2013,sothmann_magnon-driven_2012,ruokola_theory_2012,bergenfeldt_hybrid_2014,donsa_double_2014,mazza_thermoelectric_2014}. This was motivated by a number of reasons. First of all, such three-terminal setups share a number of features with mesoscopic Coulomb-drag setups that have been studied both theoretically~\cite{mortensen_coulomb_2001,goorden_two-particle_2007,goorden_cross-correlation_2008,levchenko_coulomb_2008,moldoveanu_coulomb_2009,sanchez_mesoscopic_2010,stark_coherent_2010} as well as experimentally~\cite{khrapai_double-dot_2006,mcclure_tunable_2007,zhang_noise_2007,shinkai_bidirectional_2009,laroche_positive_2011} in the last years. Such Coulomb-drag setups consist of two nearby mesoscopic conductors such as quantum dots. The first conductor is subject to a bias voltage that drives a charge current through it. Nonequilibrium charge fluctuations (noise) due to this current will induce charge fluctuations in the second, unbiased conductor. If the second conductor is intrinsically nonlinear it will rectify these induced charge fluctuations and thus exhibit a charge current without an applied bias voltage. 
% Similarly, in a three-terminal heat engine, the charge fluctuations in the first conductor that drive a charge current through the second conductor are simply thermal fluctuations due to the coupling to a hot source.
Similarly, in a three-terminal heat engine, thermal fluctuations from the hot source can get rectified and drive  a directed charge current.
Here, the nonequilibrium situation is introduced by the temperature imbalance between the hot source and the conductor. In addition, three-terminal heat engines offer the advantage of spatially separating the hot and cold reservoirs, cf. \fref{fig:harvest}. This becomes most pronounced in a recently proposed microwave-cavity heat engine where two mesoscopic conductors such as double quantum dots are connected via a superconducting microwave cavity over a typical length of a few centimeters~\cite{bergenfeldt_hybrid_2014}. The separation of hot and cold baths helps to reduce leakage heat currents that are detrimental to achieving a high efficiency of heat to work conversion. Finally, three-terminal heat engines also exhibit a crossed flow of heat and charge currents. This is useful for energy harvesting applications because it allows the hot source to be kept electrically separated from the actual energy harvester. 
% \textcolor{red}{$\rightarrow$}
Thermoelectric configurations including thermometer terminals have also been investigated~\cite{jacquet_thermoelectric_2009,caso_local_2011,sanchez_thermoelectric_2011}.
% \textcolor{red}{$\leftarrow$}

In contrast to bulk thermoelectric materials, nanoscale heat engines often operate in the nonlinear regime. Linear response theory is valid if the temperature difference on the scale of the inelastic scattering length is small compared to the average temperature~\cite{sivan_multichannel_1986,butcher_thermal_1990}. For bulk materials, the inelastic scattering length is typically much smaller than the system size, such that even for a sizable temperature bias, linear response theory is a good approximation. For nanoscale setups, system size and inelastic scattering length become comparable. Hence, nonlinear effects become important~\cite{whitney_nonlinear_2013}.
In order to properly describe nonlinear thermoelectric effects in mesoscopic structures, a nonlinear scattering matrix theory of thermoelectric transport has been developed recently~\cite{sanchez_scattering_2013,meair_scattering_2013,whitney_thermodynamic_2013}. It has been applied to study heat transport through quantum dots where, e.g., a deviation from the Wiedeman-Franz law was found~\cite{lopez_nonlinear_2013}. It was also used to investigate the magnetic-field asymmetry of nonlinear transport coefficients~\cite{hwang_magnetic-field_2013}. Experimentally, the nonlocal thermoelectric response of a ballistic four-terminal structure has been investigated~\cite{matthews_thermally_2012}. Further studies theoretically analyzed nonlinear thermoelectric transport through molecular junctions~\cite{leijnse_nonlinear_2010,hershfield_nonlinear_2013}. An important feature of nonlinear thermoelectrics is that the figure of merit $ZT$ is no longer sufficient to characterize the thermoelectric performance~\cite{whitney_nonlinear_2013}. Instead, one has to rely on quantities such as the maximal efficiency, the efficiency at maximum power~\cite{van_den_broeck_thermodynamic_2005,esposito_universality_2009,brandner_strong_2013} or the maximal efficiency at a given output power~\cite{whitney_most_2014}.

Here, we review recent theoretical work on three-terminal thermoelectrics with quantum dots. The review is organized as follows. In Sec.~\ref{sec:coulomb} we discuss systems of Coulomb-coupled conductors ranging from quantum dots in the Coulomb-blockade regime over chaotic cavities to resonant tunneling quantum dots.
The second part, Sec.~\ref{sec:boson}, focusses on quantum dot heat engines that are driven by bosonic degrees of freedom such as phonons, magnons and microwave photons.
We finish with conclusions and an outlook in Sec.~\ref{sec:concl}.

%\section{\label{sec:coulomb}Coulomb-coupled quantum dots}
\section{\label{sec:coulomb}Harvesting from electronic sources}
In the following, we discuss different types of heat engines based on quantum dots that are coupled to a hot electronic reservoir. As a first example, we will analyze a setup based on capacitively coupled quantum dots in the Coulomb-blockade regime. We then continue the discussion with a similar system based on chaotic cavities coupled to reservoirs via quantum point contacts with a large number of open transport channels. Finally, we demonstrate that heat engines based on resonant tunneling through either quantum dots or quantum wells can yield high power in combination with good efficiency.

\subsection{\label{ssec:CB}Coulomb-blockade regime}
Here, we analyze the thermoelectric properties of two capacitively coupled quantum dots in the Coulomb-blockade regime in a three-terminal geometry~\cite{sanchez_optimal_2011}. 
% \textcolor{red}{$\rightarrow$}
This model emphasizes the contribution of charge fluctuations, with heat being absorbed merely by means of the Coulomb interaction between electrons in the two different conductors.
% \textcolor{red}{$\leftarrow$}
We demonstrate how a temperature bias across the system can drive a charge current which in turn can be used to generate electrical power. Furthermore, we demonstrate that the device can act as an ideal heat to current converter that can reach Carnot efficiency. 
% \textcolor{red}{$\rightarrow$}
The bipartite nature of this setup can be exploited for other purposes such as feedback control~\cite{strasberg_thermodynamics_2013}, the detection of dynamical rates~\cite{schulenborg_detection_2014} or the investigation of information flows~\cite{horowitz_thermodynamics_2014}.
% \textcolor{red}{$\leftarrow$}

\subsubsection{Model}
\begin{figure}
	\centering\includegraphics[width=.6\columnwidth]{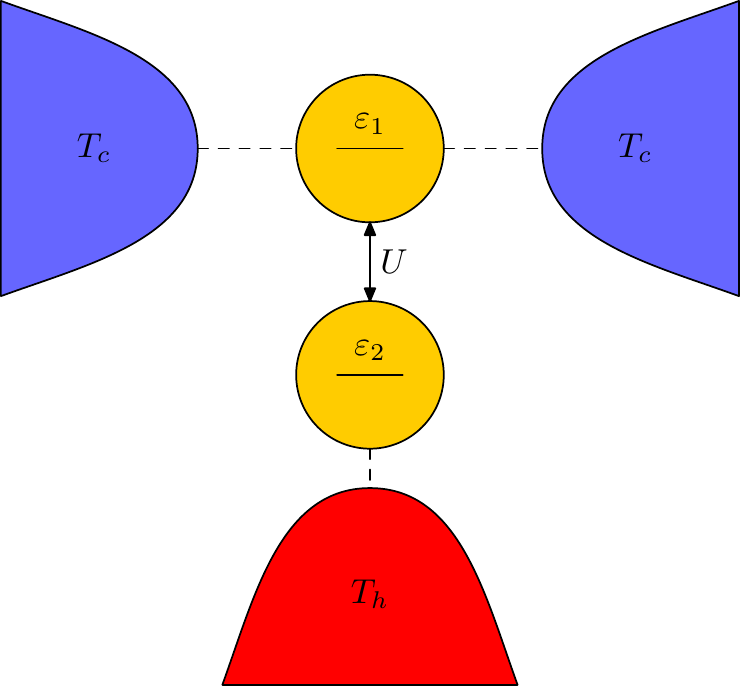}
	\caption{\label{fig:model_CB}Sketch of the Coulomb-blockade heat engine. Two single-level quantum dots (yellow) are capacitively coupled to each other. The lower gate quantum dot is connected to a single hot reservoir (red). The upper conductor dot is coupled to two cold electronic reservoirs (blue).}
\end{figure}
We consider the setup shown schematically in \fref{fig:model_CB}. It consists of two quantum dots $\alpha=1,2$ in the Coulomb-blockade regime. Each dot hosts a single level with energy $\varepsilon_\alpha$. Assuming strong onsite Coulomb interaction, the dots can host either $0$ or $1$ excess electron. In the following, we neglect the electron spin as it will only lead to a renormalization of the tunnel couplings introduced below. The two quantum dots are capacitively coupled to each other such that they can exchange energy but no particles. The strength of the coupling is characterized by the Coulomb energy $U$ that is needed to occupy both quantum dots at the same time and will parametrize the energy exchange between the two dots.
The conductor dot, $\alpha=1$, is tunnel coupled to two electronic reservoirs $r=\text{L,R}$. The reservoirs are in local thermal equilibrium and characterized by chemical potentials $\mu_r$ and temperature $T_c$. The gate dot, $\alpha=2$, is coupled to a single reservoir $\text{G}$ with chemical potential $\mu_\text{G}$ and temperature $T_h$.
In order to obtain a finite thermoelectric response through the conductor dot to a temperature difference $T_h-T_c$, we need to have energy-dependent tunnel coupling strengths. Here, we model this energy dependence by choosing the tunnel couplings $\Gamma_{sn}$ between the dot and its respective reservoir $s=\text{L,R,G}$ to depend on the number of electrons, $n$, on the \emph{other} quantum dot.
There are different ways how to realize this energy dependence in an actual experiment. First of all, the transmission through a tunnel barrier generically depends on energy of the tunneling particle. This effect is most pronounced for energies close to the barrier height. Second, quantum dots with an excited state that couples asymmetrically to the two leads and decays quickly into the ground state can mimic energy-dependent tunneling rates as well. Finally, additional quantum dots can be added between the conductor dot and its two reservoirs. These additional dots serve as energy filters that only transmit electrons at a given energy and thereby allow to achieve an effectively energy-dependent tunnel coupling.

In order to describe the system, we trace out the noninteracting electronic reservoirs. The remaining strongly interacting quantum dot degrees of freedom are then characterized by a reduced density matrix for the double quantum dot. The time evolution of the system is determined by a master equation $\dot {\vec P}=\vec W\vec P$ for the reduced density matrix elements $\vec P=(P_{00},P_{10},P_{01},P_{11})$ describing the probability to find the double dot empty $P_{00}$, occupied with one electron on the conductor $P_{10}$ or gate dot $P_{01}$ and doubly occupied $P_{11}$.
In the sequential tunneling regime $\Gamma_{ri}\ll\kB T_{h,c}$ the transition rates $\vec W$ follow from Fermi's golden rule and are given by
\begin{equation}
	\fl\vec W=
	\left(
	\begin{array}{cccc}
		-\sum_r W_{r0}^+-W_{\text{G}0}^+ & \sum_r W_{r0}^- & W_{\text{G}0}^- & 0 \\
		\sum_r W_{r0}^+ & -\sum_r W_{r0}^--W_{\text{G}1}^+ & 0 & W_{\text{G}1}^- \\
		W_{\text{G}0}^+ & 0 & -\sum_r W_{r1}^+-W_{\text{G}0}^- & \sum_r W_{r1}^- \\
		0 & W_{\text{G}1}^+ & \sum_r W_{r1}^+ & -\sum_r W_{r1}^--W_{\text{G}1}^-
	\end{array}
	\right).
\end{equation}
The transition rates $W^\pm_{sn}$ describe tunneling of electrons onto or off the dot through barrier $s$ when the other dot has $n$ electrons. Specifically, they are written as
\begin{eqnarray}
	W^\pm_{r0}=\Gamma_{r0}f^\pm_r(\varepsilon_1),\\
	W^\pm_{\text{G}0}=\Gamma_{\text{G}0}f^\pm_g(\varepsilon_2),\\
	W^\pm_{r1}=\Gamma_{r1}f^\pm_r(\varepsilon_1+U),\\
	W^\pm_{\text{G}1}=\Gamma_{\text{G}1}f^\pm_g(\varepsilon_2+U),
\end{eqnarray}
where $f_r^+(x)=1-f_r^-(x)=\{\exp[(x-eV_r)/(\kB T_r)]+1\}^{-1}$ denotes the Fermi function with temperature $T_r$ and bias voltage $V_r=\mu_r/e$.

In order to calculate the charge and heat currents as well as their fluctuations and correlations, we use a full-counting statistics approach. To this end, we multiply the transition rates with a factor $e^{iq_s\chi_s}$ where $q_s$ denotes the charge transferred through barrier $s$ in the associated tunneling event. The cumulant generating function $\mathcal S(\chi_s)$ is then given by the eigenvalue of the resulting matrix $\vec W_\chi$ that goes to zero as $\chi\to0$. The charge currents are simply given by the derivative of the cumulant generating function with respect to the counting fields, $I_s=\left.\partial \mathcal S/\partial \chi_s\right|_{\chi_s=0}$.
Similarly, for the energy currents, we introduce counting factors $e^{iE_s\xi_s}$ where $E_s$ denotes the energy transferred through barrier $s$ in the corresponding tunneling event. 
% \textcolor{red}{$\rightarrow$}
Higher-order derivatives give the correlations that will be discussed in section~\ref{sec:CBfluct}.
% \textcolor{red}{$\leftarrow$}
Finally, heat currents follow from charge and energy currents as $J_s^h=J_s^E-V_sI_s$. We remark that while charge and energy currents are conserved, $\sum_s I_s=\sum_s J_s^E=0$, heat currents in general are not conserved due to Joule heating.

\subsubsection{Heat-driven current}
\begin{figure}
	\centering\includegraphics[width=.6\columnwidth]{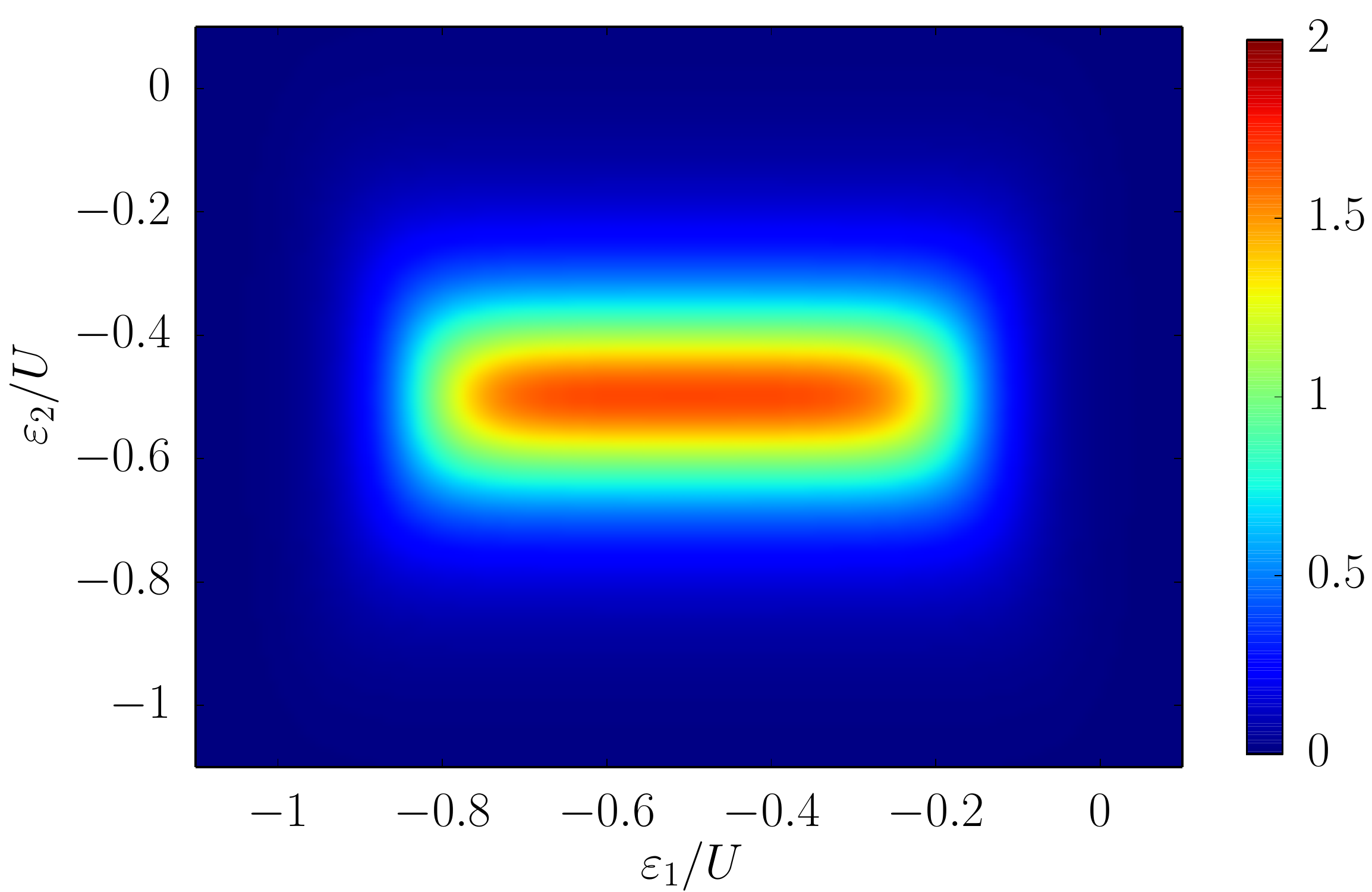}
	\caption{\label{fig:CB_current}Current at zero bias voltage in units of $10^{-3}e\Gamma$ as a function of the two level positions $\varepsilon_1$ and $\varepsilon_2$. Parameters are $\Gamma_{\text{L}0}=\Gamma_{\text{R}1}=\Gamma_{\text{G}0}=\Gamma_{\text{G}1}=\Gamma$, $\Gamma_{\text{L}1}=\Gamma_{\text{R}0}=0$, $T_c=0.4T_h$ and $U=10\kB T_h$.}
\end{figure}

In the following, we want to drive a charge current through the conductor dot simply by applying a temperature difference between the reservoirs of conductor and gate dot, i.e. without applying a bias voltage across the conductor dot. In order to achieve this goal, two requirements have to be met. First of all, the left-right symmetry of the system has to be broken. This can be achieved by choosing the tunnel couplings of the conductor dot to the left and right reservoirs different. In addition, we also have to break the particle-hole symmetry underlying the system. This is achieved by having energy-dependent tunneling rates for the conductor dot.

In the absence of a bias voltage, the charge current $I$ through the conductor dot can be simply related to the heat current $J$ flowing out of the gate dot,~\cite{sanchez_optimal_2011}
\begin{equation}\label{eq:CB_current}
	I=\frac{e}{U}\frac{\Gamma_{\text{L}0}\Gamma_{\text{R}1}-\Gamma_{\text{L}1}\Gamma_{\text{R}0}}{(\Gamma_{\text{L}0}+\Gamma_{\text{R}0})(\Gamma_{\text{L}1}+\Gamma_{\text{R}1})}J.
\end{equation}
The ratio of heat and charge currents is determined by the ratio of electron charge and charging energy. Furthermore, it depends on the ratio of the different tunnel couplings that enter the problem. The direction of the charge current can be controlled via the asymmetry of tunnel couplings as well as by the direction of the heat current, i.e. by the sign of the temperature bias $T_h-T_c$. The largest current, $I=eJ/U$, can be achieved in an optimal configuration when the conductor dot couples only to, say, the left lead when the gate dot is empty and only to the right lead, when the gate dot is occupied.

The heat-driven current as a function of the level positions of the two dots is shown in \fref{fig:CB_current}. As a function of the gate dot level, $\varepsilon_2$, the current exhibits a peak at $\varepsilon_2=-U/2$ with a width given by the temperature of the hot bath, $T_h$. As a function of the conductor dot level, $\varepsilon_1$, the current has a plateau between $\varepsilon_1=0$ and $\varepsilon_1=-U$. The borders of the plateau are smeared by the temperature of the cold bath, $T_h$.

The mechanism giving rise to the heat-driven charge current is the following. Initially, the double dot is completely empty. In a first step, an electron tunnels onto the conductor dot from, say, the left reservoir. Next, an electron tunnels onto the gate dot. This requires the charging energy $U$ which is extracted from the hot reservoir. Afterwards, the electron from the conductor dot tunnels into the right lead, thereby releasing the charging energy into the cold reservoir. In a final step, the gate dot is emptied such that the system returns to its initial state. In one such transport cycle, one quantum of energy, i.e. the charging energy, is transferred from the hot to the cold, giving rise to correlations of the charge and heat currents. 
% \textcolor{red}{$\rightarrow$}
We remark that, in the optimal configuration, one electron is transmitted from the left to the right reservoir for \emph{every} absorbed quantum of heat. This is known as the tight-coupling limit.
% \textcolor{red}{$\leftarrow$}

\subsubsection{Power and efficiency}
\begin{figure}
	\centering\includegraphics[width=.45\columnwidth]{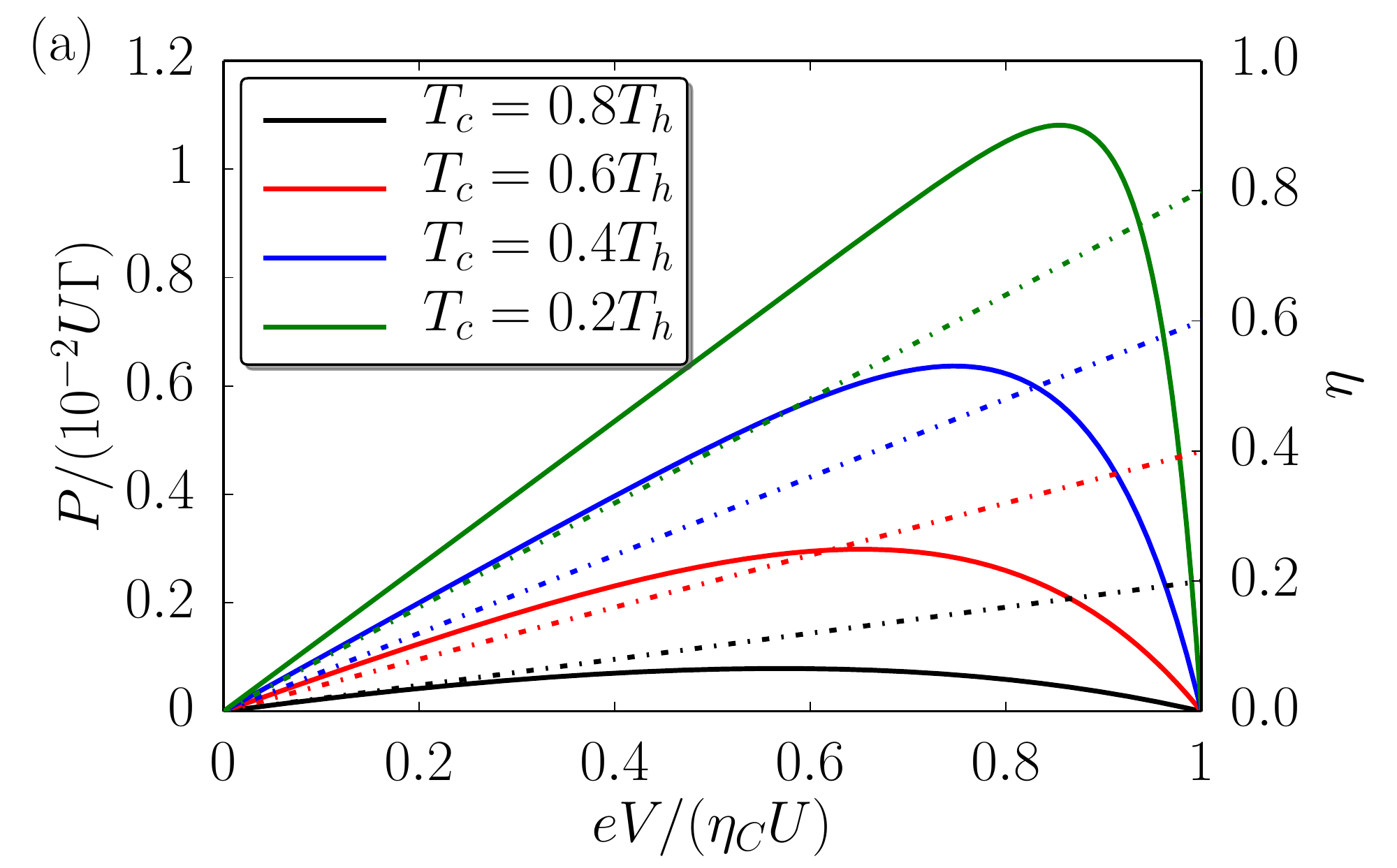}
	\includegraphics[width=.45\columnwidth]{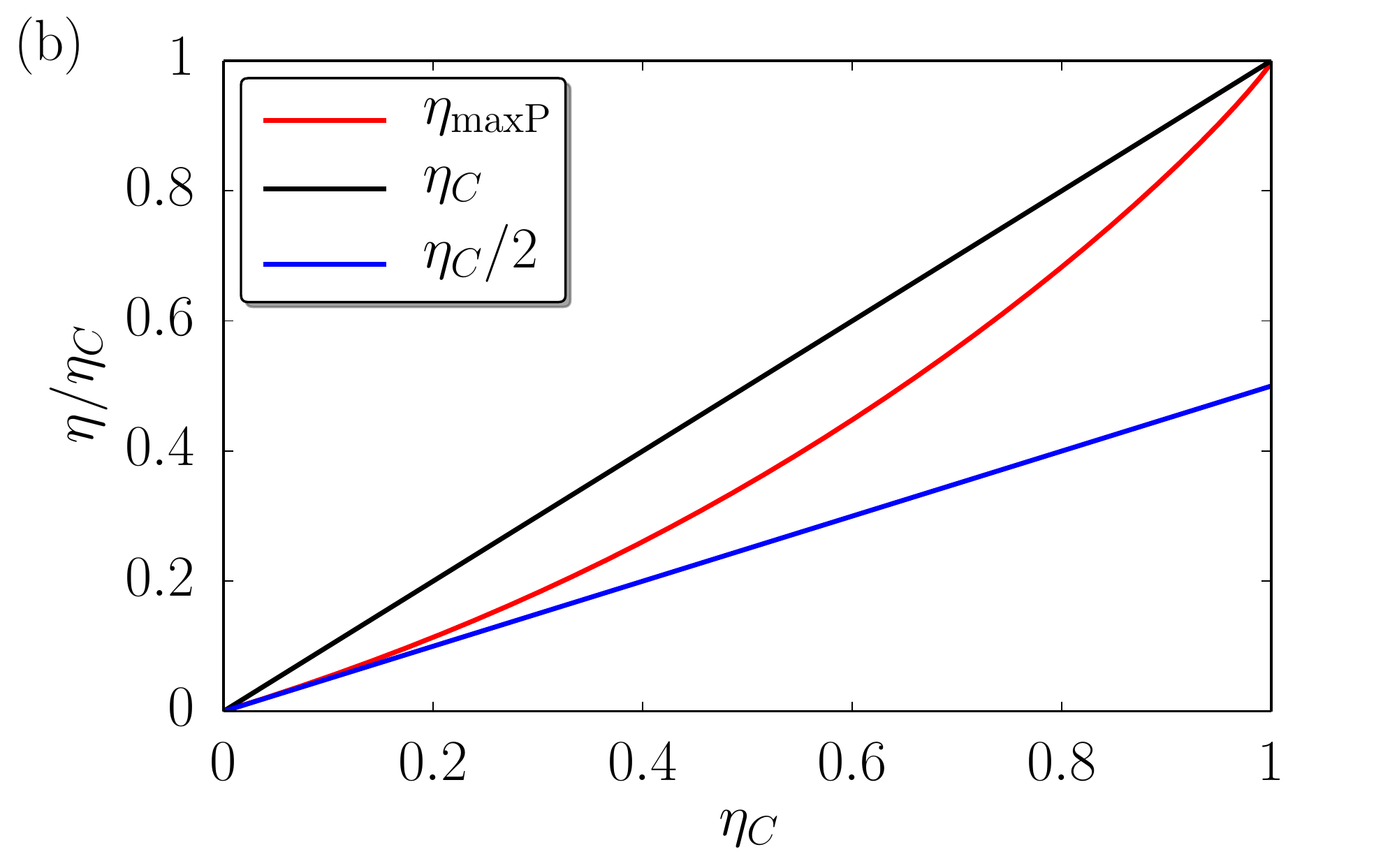}
	\caption{\label{fig:CB_power}(a) Power $P$ (solid lines) and efficiency $\eta$ (dashed lines) as a function of the applied bias voltage for different temperatures of the cold reservoir. (b) Efficiency at maximum power as a function of the Carnot efficiency. Parameters are $\Gamma_{\text{L}0}=\Gamma_{\text{R}1}=\Gamma_{\text{G}0}=\Gamma_{\text{G}1}=\Gamma$, $\Gamma_{\text{L}1}=\Gamma_{\text{R}0}=0$, $U=10T_h$, $\varepsilon_1=\varepsilon_2=-U/2$.}
\end{figure}

So far, we demonstrated that our quantum-dot heat engine can convert a heat current into a directed charge current. In a next step, we want to generate a finite output power by adding a load to the system. Hence, we apply a bias voltage $V$ across the conductor dot against which the heat-driven current can perform work. The output power is then simply given as $P=IV$. It is shown as a function of the applied bias in \fref{fig:CB_power}a). For zero bias, the output power obviously vanishes. Similarly, it vanishes at the so called stopping voltage $V_\text{stop}$ where the heat- and bias-driven currents compensate each other such that $I=0$. In between, the system reaches its maximal output power. For situations close to equilibrium, the maximum power is reached at half the stopping voltage. As the system is taken far away from equilibrium by lowering $T_c$, the point of maximum power is shifted towards larger bias voltages.

Another important quantity to characterize the thermoelectric performance of a heat engine is its efficiency of heat to work conversion. It is defined as the ratio between the output power and the input heat which in our case is given by the heat current flowing out of the hot reservoir, i.e., we have $\eta=P/J$. 
%According to \eref{eq:CB_current}, the efficiency grows linearly with the applied bias voltage,
%\begin{equation}
%	\eta=\frac{eV}{U}\frac{\Gamma_{\text{L}0}\Gamma_{\text{R}1}-\Gamma_{\text{L}1}\Gamma_{\text{R}0}}{(\Gamma_{\text{L}0}+\Gamma_{\text{R}0})(\Gamma_{\text{L}1}+\Gamma_{\text{R}1})},
%\end{equation}
%cf. also \fref{fig:CB_power}a). 
% \textcolor{red}{$\rightarrow$}
In the tight-coupling limit, the efficiency grows linearly with the applied bias voltage and we simply have $\eta=eV/U$, cf. \fref{fig:CB_power}a).
% \textcolor{red}{$\leftarrow$} 
Hence, at the stopping voltage $V_\text{stop}=U\eta_C/e$ the device reaches Carnot efficiency $\eta_C=1-T_c/T_h$ indicating that it operates as an optimal heat to current converter. However, at this point, the heat engine operates reversibly and, therefore, does not produce any output power. A more relevant quantity to consider is the efficiency at maximum power. It is shown as a function of the temperature bias in \fref{fig:CB_power}b). For a small temperature bias, it increases linearly as $\eta_C/2$ in agreement with a general thermodynamic bound for systems with time-reversal symmetry~\cite{van_den_broeck_thermodynamic_2005}. In the nonlinear regime, it grows faster than $\eta_C/2$ and even reaches $\eta_C$ for $T_c\to0$.
We remark, however, that at this point our master equation approach is no longer valid since higher order tunneling contributions that give rise to a broadening of the dot level become important.

\subsubsection{Fluctuations and noise}
\label{sec:CBfluct}
Up to now, we only discussed the average transport of heat and charge through the quantum-dot heat engine. However, at the nanoscale, fluctuations of average quantities turn out to be important~\cite{blanter_shot_2000}. Indeed, we discussed above how the generation of current depends on the correlation of charge fluctuations in the two dots.

A way to characterize these fluctuations is to look at the full counting statistics $P(N,t)$ which addresses the probability that $N$ electrons have passed through the system in a given time $t$~\cite{levitov_electron_1996,bagrets_full_2003,braggio_full_2006,flindt_counting_2008,sanchez_mesoscopic_2010}. In a quantum dot system, this quantity can actually be measured by coupling a charge sensor such as a quantum point contact or a single-electron transistor to the quantum dot~\cite{gustavsson_counting_2006,fujisawa_bidirectional_2006,flindt_universal_2009}. As the number of electrons on the dot changes, the electrical potential felt by the charge sensor changes, thus leading to a change in the current through it. This allows to measure the charge on the quantum dot in real time.

Interestingly, in the quantum-dot heat engine we can not only access the full counting statistics of charge but also of heat~\cite{sanchez_detection_2012}. 
Measuring the charge of both, the conductor and the gate dot, e.g., via a quantum point contact that couples asymmetrically to the two dots, allows us to reconstruct the counting statistics of heat from the counting statistics of charge due to the intimate relationship between heat and charge transfer in the system. This way, non-equilibrium fluctuation relations can be measured that relate the charge and heat currents~\cite{andrieux_fluctuation_2007,esposito_nonequilibrium_2009,campisi_colloquium:_2011} in terms of electron counting. This is important because in the presence of temperature gradients, fluctuation relations for charge currents~\cite{tobiska_inelastic_2005,forster_fluctuation_2008,saito_symmetry_2008} become configuration-dependent~\cite{andrieux_fluctuation_2007,krause_incomplete_2011} unless one also considers energy currents. Notably, the charge fluctuation theorem becomes universal (only depending on the thermodynamic forces) in the tight-coupling limit~\cite{sanchez_detection_2012,sanchez_erratum:_2013}.

Additional insight of the thermoelectric performance of the quantum-dot heat engine can be accessed by investigating the charge and heat current-current correlations~\cite{sanchez_correlations_2013}. The interest on electronic heat noise has appeared only very recently~\cite{averin_violation_2010,sergi_energy_2011,zhan_electronic_2011,averin_statistics_2011,crepieux_heat_2014,moskalets_floquet_2014,battista_energy_2014,utsumi_fluctuation_2014}. Already in the linear regime the charge-heat cross-correlations are related to Seebeck-like and Peltier-like coefficients by means of an extended three-terminal fluctuation-dissipation theorem. Far from equilibrium, dimensionless quantities for the auto- and cross-correlations (charge and heat Fano factors, and cross-correlation coefficient) can be defined. Divergences of the charge Fano factor can be used to measure the non-linear thermovoltage. Most interestingly, the charge-heat cross-correlations are maximal in the tight-coupling limit~\cite{sanchez_resonance_2008}, leading to Carnot-efficient configurations~\cite{sanchez_correlations_2013}
 .

%Apart from the counting statistics of heat and charge alone, one can also look at cross correlations between these quantities~\cite{sanchez_correlations_2013}. In the tight-coupling limit, one finds that heat and charge are totally correlated and that, hence, their cumulants become proportional to each other.

%------------------------------------------------------------------------------------------------------------------------------------------------------------------------
%#                                                                                                                                                                      #
%#                                                                                                                                                                      #
%#                                                                                                                                                                      #
%------------------------------------------------------------------------------------------------------------------------------------------------------------------------

\subsection{\label{ssec:open}Chaotic cavities}
As we have just discussed, capacitively coupled quantum dots in the Coulomb-blockade regime can operate as optimal heat to charge current converters that can reach Carnot efficiency. Yet, they are of limited practical use for energy harvesting applications as they provide only small currents and output power. This is a consequence of the fact that transport in these systems occurs via the tunneling of single electrons. In the following we elucidate the thermoelectric performance of a device based on chaotic cavities coupled to electronic reservoirs via quantum point contacts with a large number of open transport channels~\cite{sothmann_rectification_2012}. Our main aim is to discuss how current, power and efficiency behave as the number of open channels is changed.

\subsubsection{Model}
\begin{figure}
	\centering\includegraphics[width=.5\columnwidth]{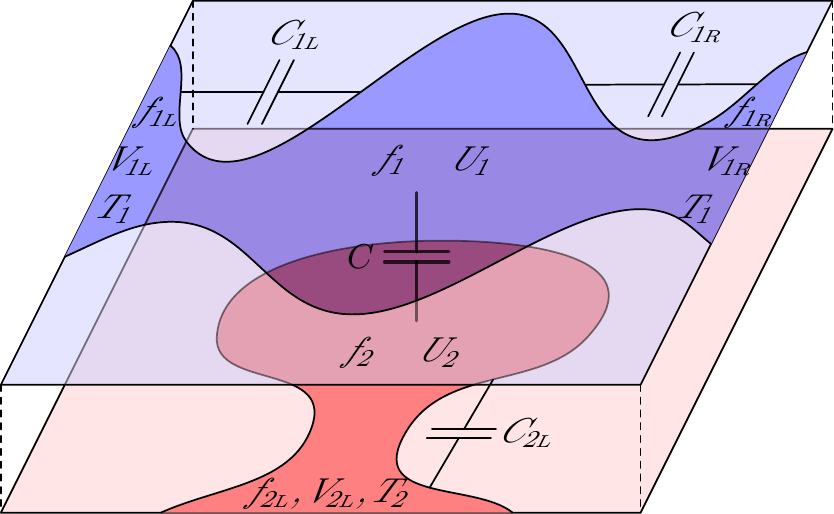}
	\caption{\label{fig:Cavity_Model}Schematic sketch of the heat engine based on chaotic cavities. The two cavities are capacitively coupled to each other. In addition, the rectifying cavity is connected to two cold electronic reservoirs at temperature $T_1$ via quantum point contacts. Similarly, the second cavity is connected to a single hot reservoir at temperature $T_2$. (Reprinted with permission from~\cite{sothmann_rectification_2012}. Copyright 2012 American Physical Society.)}
\end{figure}
We consider two open quantum dots $i=1,2$ coupled via a mutual capacitance $C$. Each cavity is connected to an electronic reservoir $r=\text{L,R}$ via a quantum point contact, cf. \fref{fig:Cavity_Model}. The reservoirs are in local thermal equilibrium and described by a Fermi function $f_r(E)=\{\exp[(E-\mu_r)/(\kB T_r)]+1\}^{-1}$ with chemical potential $\mu_r$ and temperature $T_r$. Interaction effects are captured by capacitive couplings $C_{ir}$ between cavity $i$ and reservoir $r$ that screen potential fluctuations.

We focus on the semiclassical regime where the number of open transport channels $N_{ir}$ of the quantum point contacts is large. We assume that dephasing destroys phase information but preserves energy. In this situation, the cavities are characterized by distribution functions $f_i(E)$ that depend only on energy. For later convenience, we write these distributions as
\begin{equation}
	f_i(E)=\frac{\sum_r {\cal T}_{ir} f_r(E)}{\sum_r {\cal T}_{ir}}+\delta f_i.
\end{equation}
The first term corresponds to the average of the reservoir distribution functions weighted with the transmission $\tau_{ir}$ of the corresponding QPC. The second term describes fluctuations of the distribution function that have to be determined in the following. In addition, the cavities are characterized by their potential $U_i$ and associated fluctuations $\delta U_i$. In order to obtain a finite thermoelectric response, we need to have energy-dependent transmissions ${\cal T}_{ir}$. Here, we model this energy dependence as ${\cal T}_{ir}={\cal T}_{ir}^0-e{\cal T}'_{ir}\delta U_i$. While the first term, ${\cal T}^0_{ir}$ describes the energy-independent transmissions that depend linearly on the number $N_{ir}$ of open transport channels, the second term captures changes in the transmission due to fluctuations of the cavity potential. We remark that the energy-dependent term ${\cal T}'_{ir}$ is independent of the number of open transport channels.

The distribution functions of the cavities obey a kinetic equation of the form (cf., e.g., Ref.~\cite{nagaev_frequency_2004})
\begin{equation}\label{eq:kinetic}
	e\nu_{i\text{F}}\frac{df_i}{dt}=e\nu_{i\text{F}}\frac{\partial f_i}{\partial U_i}\dot U_i+\frac{e}{h}\sum_r {\cal T}_{ir}(f_{ir}-f_i)+\delta i_\Sigma,
\end{equation}
where $\nu_{i\text{F}}$ denotes the density of states at the Fermi energy in cavity $i$. The kinetic equation describes how the charge in a given energy interval changes due to changes of the cavity potential $U_i$, in- and outgoing electron currents through the quantum point contacts as well as due to fluctuations of these currents $\delta i_\Sigma$. Here, the index $\Sigma$ indicates that we have to sum over all contacts $r$ of a given cavity $i$.

In a next step, we obtain a relation between the fluctuations of the cavity distributions $\delta f_i$ and potentials $\delta U_i$ by expressing the charge inside each cavity in terms of an integral over the distribution functions as well as in terms of the various capacitances and potentials. Using this relation between $\delta f_i$ and  $\delta U_i$, we can transform the kinetic equation~\eref{eq:kinetic} into a Langevin equation for the potential fluctuations $\delta U_i$.
Due to the nonlinearity introduced into the problem by the energy-dependent transmission, the Langevin equation has a multiplicative noise term that leads to the It\^o-Stratonovich problem in the interpretation of the stochastic integral when converting the Langevin equation into a Fokker-Planck equation~\cite{van_kampen_validity_1981,hanggi_stochastic_1982}. For the problem at hands, it turns out that only the kinetic prescription by Klimontovich~\cite{klimontovich_ito_1990} provides a meaningful solution that exhibits vanishing heat and charge currents in global equilibrium. From the Fokker-Planck equation, we can obtain the expectation values of the potential fluctuations, $\langle\delta U_i\rangle$ and $\langle\delta U_i\delta U_j\rangle$, which subsequently allow us to evaluate the charge current between cavity 1 and its contact $r$ via the standard scattering matrix expression
\begin{equation}
	I_{1r}=\frac{e}{h}\int dE {\cal T}_{1r}(f_{1r}-f_{1})+\delta I_r.
\end{equation}

\subsubsection{Current, Power and Efficiency}
We first focus on the situation where a finite temperature bias $T_1-T_2$ is applied across the system while there is no bias voltage applied across cavity 1, i.e. $V_{1\text{L}}=V_{1\text{R}}$. For the charge current through cavity 1 we find to lowest order in the energy-dependent transmission the compact expression
\begin{equation}\label{eq:cavity_current}
	\langle I_\text{1L}\rangle=\frac{\Lambda}{\tau_{RC}} \kB(T_1-T_2).
\end{equation}
The charge current depends on the asymmetry parameter
\begin{equation}\label{eq:Lambda}
	\Lambda =\frac{G'_{1\text{L}}G_{1\text{R}}-G'_{1\text{R}}G_{1\text{L}}}{G_{1\Sigma}^2},
\end{equation}
with $G_{ir}=(e^2/h) {\cal T}^0_{ir}$, $G_{i\Sigma}=G_{i\text{L}}+G_{i\text{R}}$ and $G'_{ir}=(e^3/h) {\cal T}'_{ir}$. It characterizes both the breaking of left-right symmetry as well as the breaking of particle-hole symmetry due to energy-dependent transmissions. From Eq.~\eref{eq:Lambda}, we conclude that in order to have a finite charge current driven by the temperature bias, we need to break both symmetries simultaneously. 
% \textcolor{red}{$\rightarrow$}
We remark that a similar antisymmetric combination of energy-dependent transmissions was found in the expression for the charge current through Coulomb-blockaded dots,~\eref{eq:CB_current}, as there we also have to break left-right and particle-hole symmetry at the same time.
% \textcolor{red}{$\leftarrow$}
In addition, the current also depends on the $RC$ time of the cavity, $\tau_{RC}=C_\text{eff}/G_\text{eff}$. Here, $G_\text{eff}=G_{1\Sigma}G_{2\Sigma}/(G_{1\Sigma}+G_{2\Sigma})$ denotes the effective conductance of the double cavity while $C_\text{eff}$ is an effective capacitance that characterizes the coupling between the two cavities. For weakly coupled cavities, it grows as $C^{-2}$.

We now turn to the discussion of how the current depends on the number of open transport channels. As the energy-dependent part of the transmission does not scale with $N_{ir}$, we find that the current is independent of the number of open channels. For realistic parameters~\cite{van_wees_quantum_1991,patel_properties_1991,rossler_transport_2011} with $C_\text{eff}=\unit[10]{fF}$, $G'=(e^2/h)\unit[]{(mV)}^{-1}$ and a temperature bias of \unit[1]{K}, we obtain $I\sim\unit[0.1]{nA}$. Hence, the current for the cavity heat engine is about two orders of magnitude larger than for the heat engine operating in the Coulomb-blockade regime. The reason for this enhanced current is simply the difference between transport through a tunnel barrier and a fully open quantum channel.

In order to generate a finite output power, we have to apply a bias voltage $V_{1\text{L}}-V_{1\text{R}}$ against which the heat-driven charge current through cavity 1 can perform work. The output power is then simply given by $P=\langle I_\text{1L} \rangle(V_{1\text{L}}-V_{1\text{R}})$. It vanishes when no bias voltage is applied. Similarly, it also vanishes at the stopping voltage $V_\text{stop}= \Lambda\, \kB(T_1-T_2)/ (G_1 \tau_{RC})$ (here, $G_1=G_{1\text{L}}G_{1\text{R}}/G_{1\Sigma}$ denotes the  conductance of cavity 1) where heat- and bias-driven charge currents exactly compensate each other such that $\langle I_\text{1L} \rangle=0$. The maximal power is generated at half the stopping voltage and is given by
\begin{equation}
	P_\text{max}=\frac{\Lambda^2}{4G_1\tau_{RC}^2}\left[\kB(T_1-T_2)\right]^2.
\end{equation}
Surprisingly, the output power drops inversely as the number of open transport channels. 
The reason for this lies in the fact that the energy-dependent transmission is independent of the number of open transport channels. While the heat-driven current thus is independent of the channel number as well, the bias driven current linearly scales with the channel number. Hence, the more transport channels are open, the smaller the stopping voltage and, hence, the smaller the output power that can be achieved.

We now turn to the discussion of the efficiency $\eta$ given by the ratio of output power to input heat. The input heat is given by the heat current flowing between the cavities. To leading order in the nonlinearity, it is given by
\begin{equation}
	J_H=\frac{1}{\tau_{RC}}\kB(T_2-T_1),
\end{equation}
i.e. there is a finite heat current even in the absence of energy-dependent transmissions. As the heat current is independent of the applied bias voltage, we find that the maximal efficiency and the efficiency at maximum power coincide and are given by
\begin{equation}
	\eta_\text{max}=\frac{\Lambda^2}{4G_1\tau_{RC}}\kB(T_2-T_1).
\end{equation}
Analyzing the scaling behaviour of the efficiency with the number of transport channels, we find that it drops inversely to the square of the number of open channels. This faster decrease as compared to the output power is due to the proportionality of the heat current to the number of transport channels. For realistic parameters, we find that for a few open channels an output power of a few fW can be generated. At the same time, the efficiency reaches at most a few percent of the Carnot efficiency for a device operating a liquid helium temperatures.

%------------------------------------------------------------------------------------------------------------------------------------------------------------------------
%#                                                                                                                                                                      #
%#                                                                                                                                                                      #
%#                                                                                                                                                                      #
%------------------------------------------------------------------------------------------------------------------------------------------------------------------------

\subsection{\label{ssec:resonant}Resonant-tunneling quantum dots}
\begin{figure}
	\centering
	\includegraphics[width=.6\columnwidth]{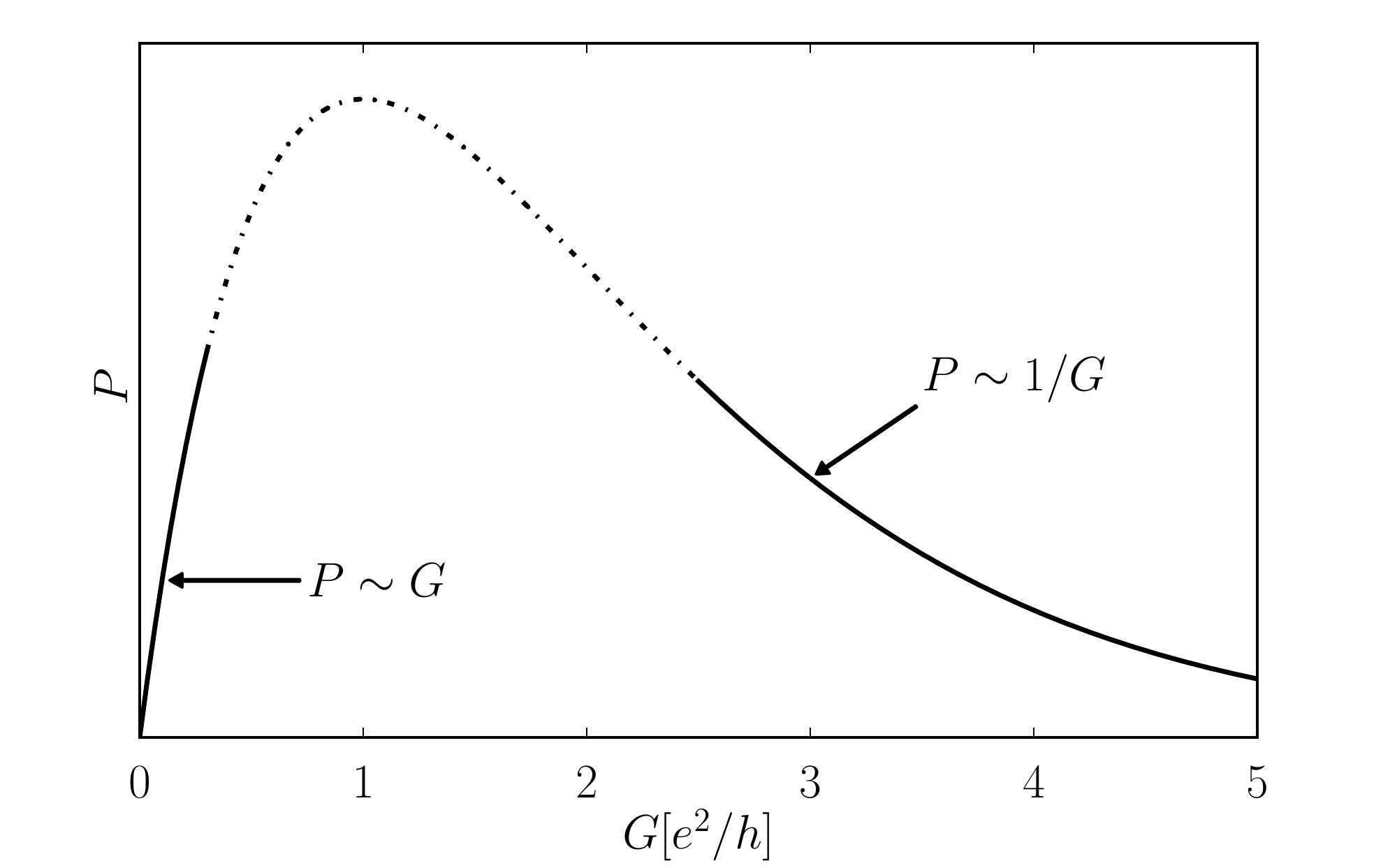}
	\caption{\label{fig:PowerScheme}Schematic dependence of the output power $P$ of a heat engine on its conductance $G$. In the Coulomb-blockade regime, power grows linearly with conductance until the Coulomb-blockade regime is left. For large conductances, power decays as the inverse conductance. In between, the maximal output power is reached for a heat engine that operates with a single open transport channel.}
\end{figure}
So far, we discussed heat engines based on quantum dots in the Coulomb-blockade regime and open quantum dots. We found that both deliver a rather small power. In the Coulomb-blockade regime, power grows linearly with the conductance but is limited by the fact that transport occurs via the tunneling of single electrons. For chaotic cavities, the power drops inversely with the conductance because the relative importance of the energy-dependent transmission goes down as more and more transport channels open up, cf. the schematic sketch in \fref{fig:PowerScheme}. The maximal output power can therefore be expected for a heat engine based on transport through a single quantum channel. A paradigmatic realization of such single channel transport is given by resonant tunneling through quantum dots that we analyze below~\cite{jordan_powerful_2013}. Similar setups have been considered in their dual role as an electronic refrigerator~\cite{edwards_quantum-dot_1993,edwards_cryogenic_1995} and successfully been used to cool a micrometer-sized island from \unit[280]{mK} to \unit[190]{mK}~\cite{prance_electronic_2009}.

\subsubsection{Setup}
\begin{figure}
	\centering\includegraphics[width=.7\columnwidth]{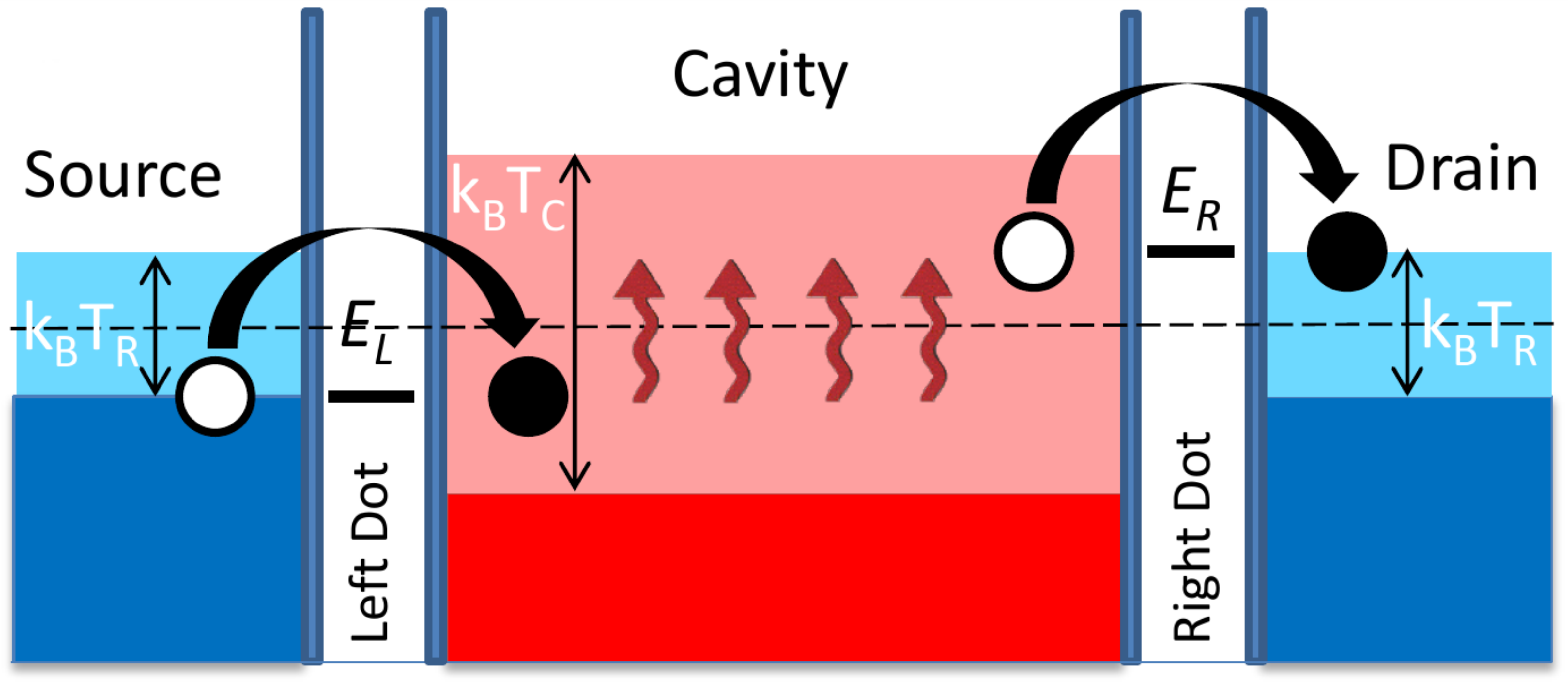}
	\caption{\label{fig:ResonantDot_Model}Heat engine based on resonant tunneling through quantum dots. A central cavity (red) at temperature $T_\text{C}$ is coupled via quantum dots to electronic reservoirs (blue) at temperature $T_\text{R}$. Light areas in the cavity and reservoirs indicate the thermal broadening of the Fermi distribution. 
% 	In a) the device is shown in a rectification configuration where the heat-driven current flows without an applied bias voltage. In b) the Carnot efficient stoppping configuration is shown where the heat-driven charge current is exactly compensated by a charge current driven by an externally applied bias voltage $V$. 
	(Reprinted with permission from~\cite{jordan_powerful_2013}. Copyright 2013 American Physical Society.)}
\end{figure}
We consider a setup consisting of a central cavity connected to two electrodes via quantum dots, cf. \fref{fig:ResonantDot_Model}. The electrodes are assumed to be in local thermal equilibrium characterized by a Fermi distribution 
% $f(E)=\{\exp[(E-\mu_r)/(\kB T_\text{R} )]+1\}^{-1}$ HAS BEEN DEFINED ABBOVE
with temperature $T_\text{R}$ and chemical potential $\mu_{L,R}$. Each quantum dot has a single resonant level relevant for transport. The levels are characterized by their width $\gamma$ which we assume to be the same for both dots in the following and their level position $E_{L,R}$. The difference of level positions $\Delta E=E_R-E_L$ characterizes the energy an electron can gain in passing through the cavity. It is different from the level spacing within each dot and can be tuned by applying a gate voltage. The central cavity is in thermal contact with a hot bath. 
% \textcolor{red}{$\rightarrow$}
This coupling is treated as a third terminal that injects a heat current $J$ but no charge in the conductor, as in \fref{fig:harvest}. For our purposes here, the nature of the hot source needs not to be further specified.
% \textcolor{red}{$\leftarrow$}
In contrast to the previous model, where the cavity was assumed out of equilibrium, here we make the simple assumption that fast relaxation processes via electron-electron and electron-phonon scattering give rise to a Fermi distribution of electrons inside the cavity with temperature $T_\text{C}$ and chemical potential $\mu_C$. The cavity temperature and chemical potential are determined by the conservation of charge and energy flowing into the cavity $I_L+I_R=0$ and $J_L+J_R+J=0$ where 
\begin{equation}\label{eq:currentC_res}
	I_j = \frac{2e}{h} \int dE\, {\cal T}_j(E) [f_j(E)-f_C(E)]
\end{equation}
denotes the charge current flowing between the cavity and reservoir $j=L,R$ while 
\begin{equation}\label{eq:currentE_res}
	J_j = \frac{2}{h} \int dE\, {\cal T}_j(E) E [f_j  {-} f_C]
\end{equation}
is the energy current flowing between the cavity and reservoir $j$. In the above expressions for the currents
\begin{equation}
	{\cal T}_j(E){=} \frac{\gamma^2}{(E{-}E_j)^2 {+} \gamma^2},
\end{equation}
denotes the transmission function of the quantum dot levels~\cite{buttiker_coherent_1988}. It is a Lorentzian with width $\gamma$ centered around the level position $E_{L,R}$. The charge conservation condition can be satisfied by placing the cavity potential symmetrically between the resonant levels of the quantum dots $\mu_C=(E_L+E_R)/2$ and putting the chemical potentials of the reservoirs symmetrically with respect to $\mu_C$, $\mu_r=\mu_C\pm\mu/2$.

\subsubsection{\label{ssec:QDResult}Results}
We first discuss the regime $\gamma\ll\kB T_\text{R},\kB T_\text{C}$ in which an analytical solution of the problem can be obtained that provides us with an intuitive understanding of the underlying physics. Afterwards, we will discuss the regime $\gamma\sim\kB T_\text{R},\kB T_\text{C}$ which we numerically find to yield the largest output power.

With narrow resonances, an electron that enters the cavity from the left lead with energy $E_L$ has to gain the precise amount of energy $\Delta E$ in order to be able to leave the cavity into the right lead. In the steady state, rectified charge current $I=I_L$ and heat current $J$ are therefore proportional to each other with the proportionality constant being given by the ratio between electron charge and difference of level positions,
\begin{equation}
	I = \frac{e}{\Delta E}J,
\end{equation}
again defining a tight-coupling limit.
We now assume that no bias voltage is applied between the two electrodes. We inject a heat current $J$ in such a way as to keep the cavity at a given temperature $T_\text{C}$. As a result, we find that the charge current
\begin{equation}\label{eq:current_res}
	I = e J/\Delta E \approx \frac{e \gamma \Delta E}{4 h}[(k_B T_\text{R})^{-1} - (k_B T_\text{C})^{-1}],
\end{equation}
flows in response to the temperature difference between the hot cavity and the cold electrodes. The above expression is valid in the limit where $k_B T_\text{R}, k_B T_\text{C} \gg \Delta E$. From \eref{eq:current_res} we conclude that the current and, hence, the output power grow both with the level width as well as with the difference of level positions until these quantities exceed the temperature.

The output power that can be generated against an externally applied bias voltage $V=\mu/e$ is given by
\begin{equation}
	P=\frac{\gamma}{4h\kBT_\text{R}}\mu(\mu_\text{stop}-\mu),
\end{equation}
for temperatures larger than $eV$ and $\Delta E$. Here, $\mu_\text{stop}=\Delta E(1-T_\text{R}/T_\text{C})$ denotes the stopping voltage at which heat- and bias-driven current compensate each other. At the stopping voltage, the device operates reversibly and reaches Carnot efficiency. At half the stopping voltage, the output power becomes maximal with
\begin{equation}\label{eq:power_res}
	P_{\rm max} \approx  \frac{\gamma \Delta E^2 \eta_C^2}{16 h k_B T_R}.
\end{equation}
At this point, the efficiency of heat to work conversion is given by $\eta_C/2$ in agreement with general thermodynamic bounds for time-reversal symmetric systems~\cite{van_den_broeck_thermodynamic_2005,esposito_universality_2009,benenti_thermodynamic_2011}.

\begin{figure}
	\centering\includegraphics[width=.75\columnwidth]{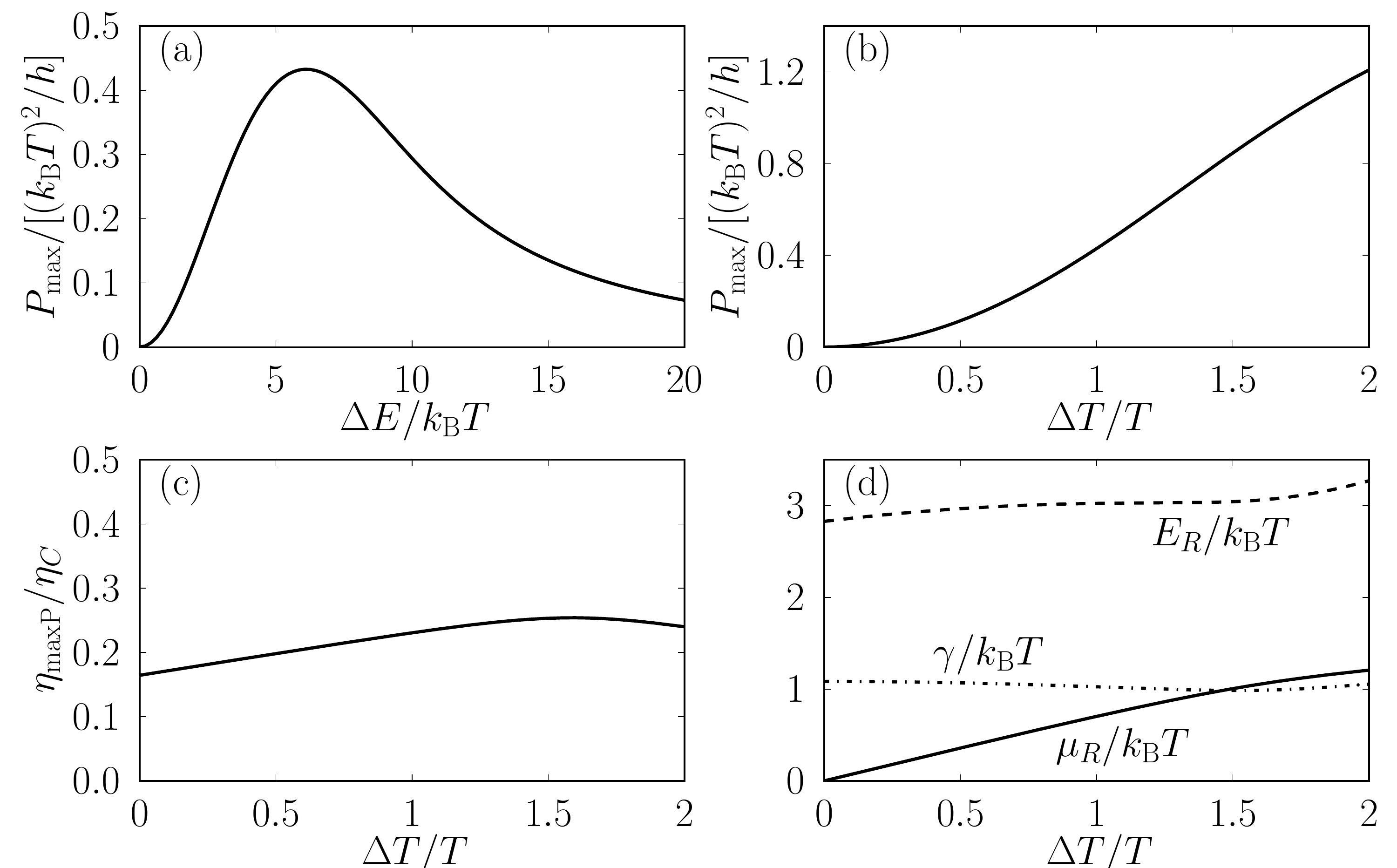}
	\caption{\label{fig:Analysis_ResDot}(a) Maximal power as a function of the energy splitting $\Delta E$ for $\Delta T=T$ and level position and width optimized for maximum power. (b) Maximum power and (c) efficiency at maximum power as a function of $\Delta T/T$. The corresponding optimal values of $E_R=\Delta E/2$, $\gamma$ and $\mu_R$ are shown in (d). (Reprinted with permission from~\cite{jordan_powerful_2013}. Copyright 2013 American Physical Society.)}
\end{figure}

We now turn to the situation of arbitrary level width and position. In this case, the integrals in \eref{eq:currentC_res} and \eref{eq:currentE_res} have to be evaluated numerically. The main results of our analysis are summarized in \fref{fig:Analysis_ResDot} where we defined the average temperature $T=(T_\text{C}+T_\text{R})/2$ and the temperature bias $\Delta T=T_\text{C}-T_\text{R}$. \Fref{fig:Analysis_ResDot}(a) shows the output power as a function of the energy difference $\Delta E$. For small $\Delta E$ the power grows quadratically with $\Delta E$ in agreement with the analytical result~\eref{eq:power_res}. 
After reaching a maximum at $\Delta E\approx 6\kBT$ it decays exponentially for large $\Delta E$ due to the small number of electrons available at very high energies. \Fref{fig:Analysis_ResDot}(b) shows the maximal output power we can obtain when optimizing bias voltage, level positions and level width at the same time. The corresponding optimal parameters are depicted in \Fref{fig:Analysis_ResDot}(d). While the optimal bias voltage grows linearly with the applied temperature bias, the optimal level position and width are nearly independent of it and given by $\Delta E\approx 6\kBT$ and $\gamma\approx\kBT$. The maximal power is given by $P_\text{max}\sim 0.4 (\kB\Delta T)^2/h$ which amounts to about \unit[0.1]{pW} for temperature bias of \unit[1]{K}. At the same time, the efficiency at maximum power is nearly independent of the temperature bias and given by about $0.2\eta_C$. Compared to the Coulomb-blockade regime, we therefore loose a factor of two in efficiency. This is, however, more than compensated by the two orders of magnitude gain in output power.  The reason for this dramatic increase in power is the combination of a strong energy dependence of the transmission functions in combination with a large number of electrons that can pass through a fully open quantum channel. We estimate that an area of $\unit[1]{cm^2}$ covered with  nanoengines that each have an area of $\unit[100]{nm^2}$ can produce an output power of \unit[10]{W} for a temperature bias of \unit[10]{K}.

\begin{figure}
	\centering\includegraphics[width=.5\columnwidth]{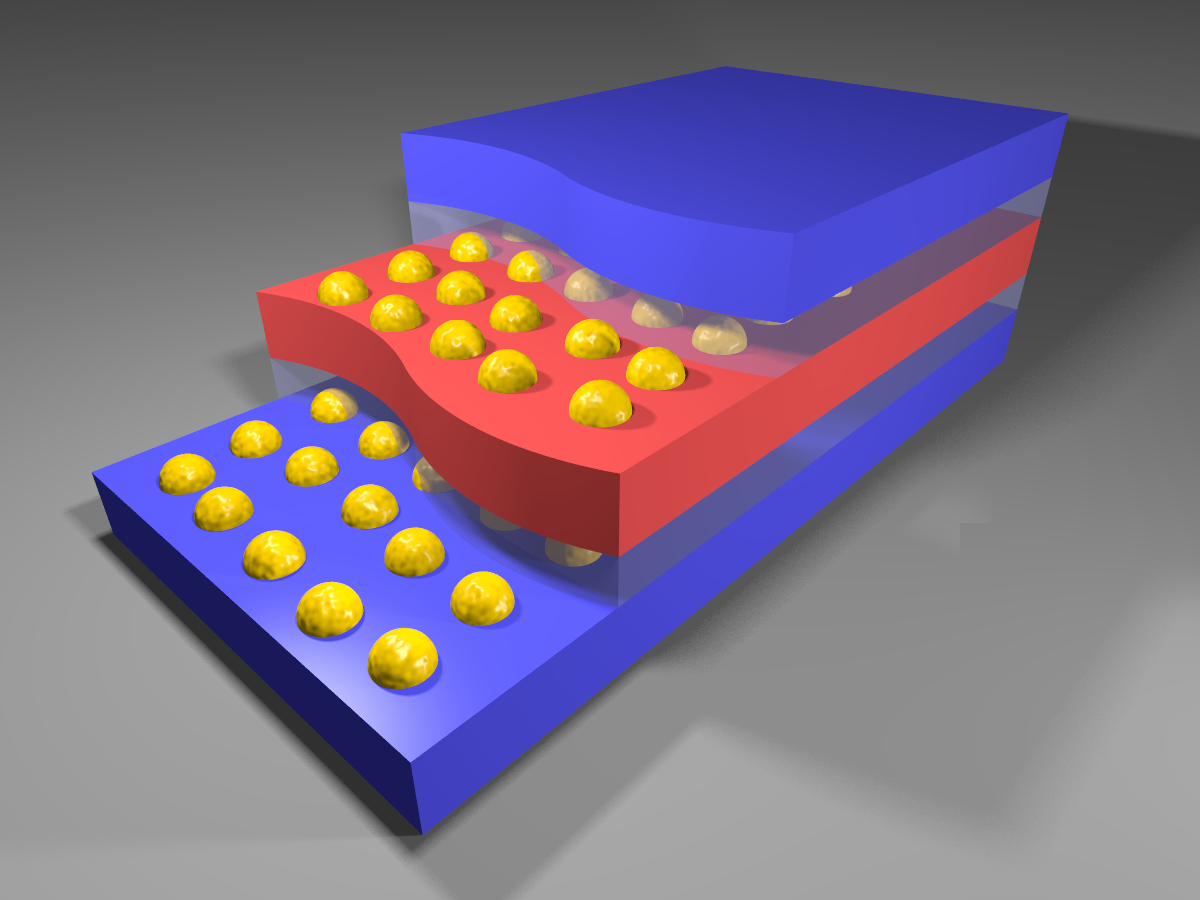}
	\caption{\label{fig:swiss_cheese}Swiss-cheese sandwich heat engine. A large central cavity (red) is connected via layers of self-assembled quantum dots (yellow) embedded into an insulating matrix (transparent) to two cold electronic reservoirs (blue).}
\end{figure}
Such a high packing density can be achieved by a strongly parallelized setup shown in \fref{fig:swiss_cheese}. Here, the electrodes are connected to a large central cavity via layers of self-assembled quantum dots embedded into an insulating matrix. Electrons tunnel through the quantum dots like through holes in a slice of Swiss cheese driven by the thermal bias. Importantly, the positions of the dots in the two layers do not have to match.
Apparently, the Swiss-cheese sandwich heat engine outperforms a heat engine based on chaotic cavities discussed above. The reason for this lies in the fact that here we put many optimized channels in parallel while for open quantum dots there are many channels in parallel out of which only a single one is relevant for thermoelectric purposes. We remark that the heat engine based on self-assembled quantum dots offers the additional advantage that the irregular nature of the quantum dots layers can help to increase scattering of phonons at the interface and thereby reduce unwanted leakage heat currents between the hot cavity and the cold reservoirs. 

\begin{figure}
	\centering\includegraphics[width=.5\columnwidth]{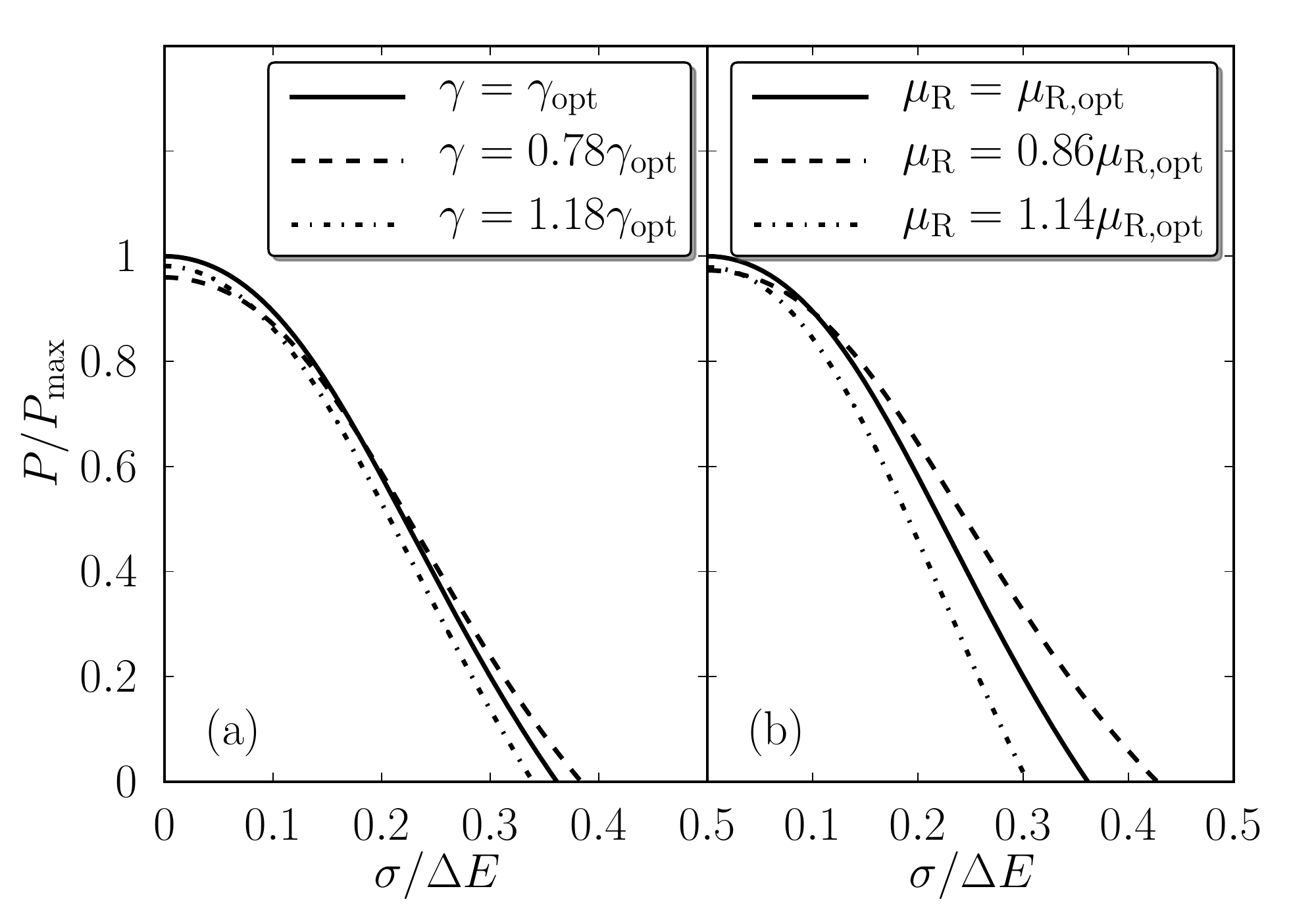}
	\caption{\label{fig:fluctuations}Output power as a function of the width $\sigma$ of the Gaussian distribution of level positions. In (a) the influence of level width that differ from the optimal value are shown. Similarly, in (b) the influence of deviations from the optimal bias voltage are shown. (Reprinted with permission from~\cite{jordan_powerful_2013}. Copyright 2013 American Physical Society.)}
\end{figure}
So far, we considered the ideal situation where all dots have the same, optimal properties. In a real sample, there will be fluctuations of level positions from dot to dot. In order to investigate in how far these imperfections deteriorate the performance of the heat engine, we assume that the level positions are distributed according to a Gaussian with width $\sigma$ centered around the average level position $E_{L,R}$. The output power as a function of the distribution width is shown in \fref{fig:fluctuations}. As expected the power drops down as the width of the distribution increases. Importantly, even for a spread of 10\% the power decreases only to 90\% of its optimal value, i.e. the proposal can tolerate a certain degree of imperfections. Interestingly, for a given degree of fluctuations, choosing nonoptimal values of the level width or bias voltage can even increase the output power.

%------------------------------------------------------------------------------------------------------------------------------------------------------------------------
%#                                                                                                                                                                      #
%#                                                                                                                                                                      #
%#                                                                                                                                                                      #
%------------------------------------------------------------------------------------------------------------------------------------------------------------------------

\subsection{\label{ssec:well}Quantum wells}
In the previous section we analysed a heat engine based on resonant tunneling through quantum dots that could be scaled up using self-assembled quantum dots in order to deliver a macroscopic output power. In the following, we discuss a related proposal in which the quantum dot layers are replaced by quantum wells~\cite{sothmann_powerful_2013}. The possibility to create high thermoelectric figures of merit by using ridged quantum wells has been pointed out~\cite{tavkhelidze_large_2009}. Furthermore, it was shown that quantum well structures can be used for refrigeration~\cite{nian_solid-state_2014}. Multilayered thermionic devices which have been proposed for refrigeration purposes are similar in design but differ in the role of the resonance~\cite{mahan_multilayer_1998}.

Compared to heat engines based on resonant tunneling through quantum dots, heat engines based on quantum wells offer a number of potential advantages. First of all, electrons inside a quantum well have transverse degrees of freedom. This gives rise to a larger phase space for tunneling electrons. Hence, quantum wells potentially allow for larger currents and output powers. Second, fabricating a homogenous quantum well can be easier than growing a layer of quantum dots with identical properties (even though we have seen above that certain fluctuations in the dot properties can be tolerated). Finally, narrow quantum wells exhibit subband spacings of the order of several hundreds of meV~\cite{chang_resonant_1974,bonnefoi_resonant_1985} which makes them ideal candidates for room temperature applications including spin based thermoelectrics~\cite{nicolau_spin_2014}.
Another important difference between quantum dots and wells is that the latter transmit any electron with an energy larger than the subband threshold. Therefore, quantum wells are much less efficient energy filters which can potentially degrade their thermoelectric properties. 

\subsubsection{Model}
\begin{figure}
	\centering\includegraphics[width=.5\columnwidth]{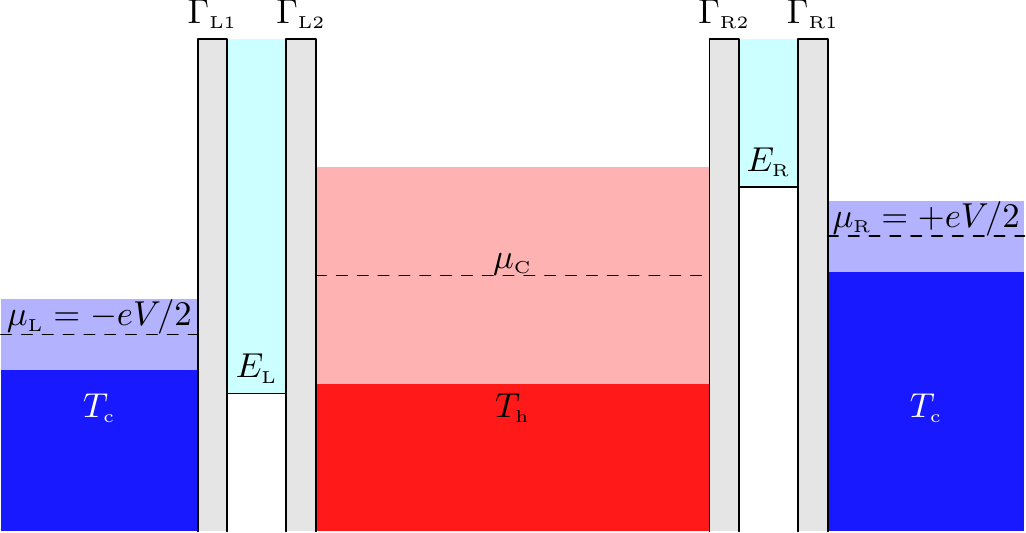}
	\caption{\label{fig:well_model}Schematic of the quantum-well heat engine. A cavity at temperature $T_h$ (red) is coupled to electronic reservoirs at temperature $T_c$ (blue). The coupling is established via quantum wells with threshold energies $E_{L,R}$. The light shading inside the quantum wells indicates that electrons with any energy larger than the threshold energy can pass through the quantum well. (Adapted with permission from~\cite{sothmann_powerful_2013}. Copyright 2013 IOP Publishing.)}
\end{figure}
We consider a setup similar to the one discussed in section~\ref{ssec:resonant}. It consists of a central cavity in local thermal equilibrium with temperature $T_h$ and chemical potential $\mu_C$. It is coupled to cold electronic reservoirs with temperature $T_c$ and chemical potentials $\mu_r$ via quantum wells, cf. \fref{fig:well_model}. We assume that the quantum wells are noninteracting such that charging effects can be neglected. It is an interesting topic of future research to investigate the influence of interactions on the thermoelectric properties of quantum-well heat engines.

The cavity potential and temperature are determined by the conservation of charge and energy, $I_\text{L}+I_\text{R}=0$ and $J_\text{L}^\text{E}+J_\text{R}^\text{E}+J=0$, where $I_r$ and $J^\text{E}_r$ denote the charge and energy current flowing between the cavity and reservoir $r$, respectively. In addition, we have the heat current $J$ that is injected into the cavity from a hot thermal bath and later on serves as a drive for the heat engine.
 
The charge and energy currents can be derived within a scattering matrix approach as~\cite{blanter_transition_1999},
\begin{equation}\label{eq:I}
	I_r=\frac{e\nu_2\mathcal A}{2\pi\hbar}\int dE_\perp dE_z {\cal T}_r(E_z)\left[f_r(E_z+E_\perp)-f_\text{C}(E_z+E_\perp)\right],
\end{equation}
and
\begin{equation}\label{eq:JE}
	J^\text{E}_{r}=\frac{\nu_2\mathcal A}{2\pi\hbar}\int dE_\perp dE_z (E_z+E_\perp){\cal T}_r(E_z)\left[f_r(E_z+E_\perp)-f_\text{C}(E_z+E_\perp)\right].
\end{equation}
In these expressions, $\nu_2=m_*/(\pi\hbar^2)$ denotes the density of states of the two-dimensional electron gas inside the quantum well with an effective electron mass $m_*$. $\mathcal A$ is the surface area of the quantum well. The energies associated with the electron motion in the quantum well plane and perpendicular to it are $E_z$ and $E_\perp$, respectively. The transmission of the quantum wells is given by~\cite{buttiker_coherent_1988}
\begin{equation}
	{\cal T}_r(E)=\frac{\Gamma_{r1}(E)\Gamma_{r2}(E)}{(E-E_{nr})^2+[\Gamma_{r1}(E)+\Gamma_{r2}(E)]^2/4}.
\end{equation}
with the (energy-dependent) coupling strengths $\Gamma_{r1}$ and $\Gamma_{r2}$ between the quantum well and reservoir $r$ or the cavity, respectively. The energies $E_{rn}$ denote the subband thresholds at which a new transport channel through the quantum well opens up. 
For the following discussion, we consider the limit that the quantum wells are only weakly coupled to the reservoirs and the cavity, $\Gamma_{r1},\Gamma_{r2}\ll\kB T_c\kB T_h$. In addition, we restrict ourselves to the case where only the lowest subband is relevant for transport. Due to the large level spacing of narrow quantum wells, this is a reasonable approximation. A discussion of thermoelectric transport through a single quantum well that takes into account higher subbands as well can be found in Ref.~\cite{agarwal_power_2014}. Under the above assumptions, the transmission function of the quantum wells simplifies to a delta peak, ${\cal T}_r(E)=2\pi\Gamma_{r1}\Gamma_{r2}/(\Gamma_{r1}+\Gamma_{r2})\delta(E_z-E_{1r})$. In this limit, the integrals in \eref{eq:I} and \eref{eq:JE} can be solved analytically and we obtain for the charge and energy currents
\begin{equation}
	I_r=\frac{e\nu_2\mathcal A}{\hbar}\frac{\Gamma_{r1}\Gamma_{r2}}{\Gamma_{r1}+\Gamma_{r2}}\left[\kB T_\text{c}K_1\left(\frac{\mu_r-E_r}{\kB T_\text{c}}\right)-\kB T_\text{h} K_1\left(\frac{\mu_\text{C}-E_r}{\kB T_\text{h}}\right)\right],
\end{equation}
and
\begin{equation}\label{eq:JEana}
	\fl J^\text{E}_r=\frac{E_{r}}{e}I_r
	+\frac{\nu_2\mathcal A}{\hbar}\frac{\Gamma_{r1}\Gamma_{r2}}{\Gamma_{r1}+\Gamma_{r2}}\left[(\kB T_\text{c})^2K_2\left(\frac{\mu_r-E_r}{\kB T_\text{c}}\right)-(\kB T_\text{h})^2K_2\left(\frac{\mu_\text{C}-E_r}{\kB T_\text{h}}\right)\right],
\end{equation}
respectively. For simplicity, we denoted the energy of the single relevant subband in each quantum well as $E_r$. Furthermore, we introduced the integrals $K_1(x)=\int_0^\infty dt(1+e^{t-x})^{-1}=\log(1+e^x)$ and $K_2(x)=\int_0^\infty dt t(1+e^{t-x})^{-1}=-{\rm Li}_2(-e^x)$ with the dilogarithm ${\rm Li}_2(z)=\sum_{k=1}^\infty \frac{z^k}{k^2}$. From \eref{eq:JEana} we find that the energy current consists of two different contributions. The first one is proportional to the charge current while the second term breaks this proportionality. It arises due to the transverse degrees of freedom and is absent in the case of quantum dots with sharp levels, cf. Sec.~\ref{ssec:QDResult}.

\subsubsection{Linear response}
\begin{figure}
	\centering\includegraphics[width=.49\columnwidth]{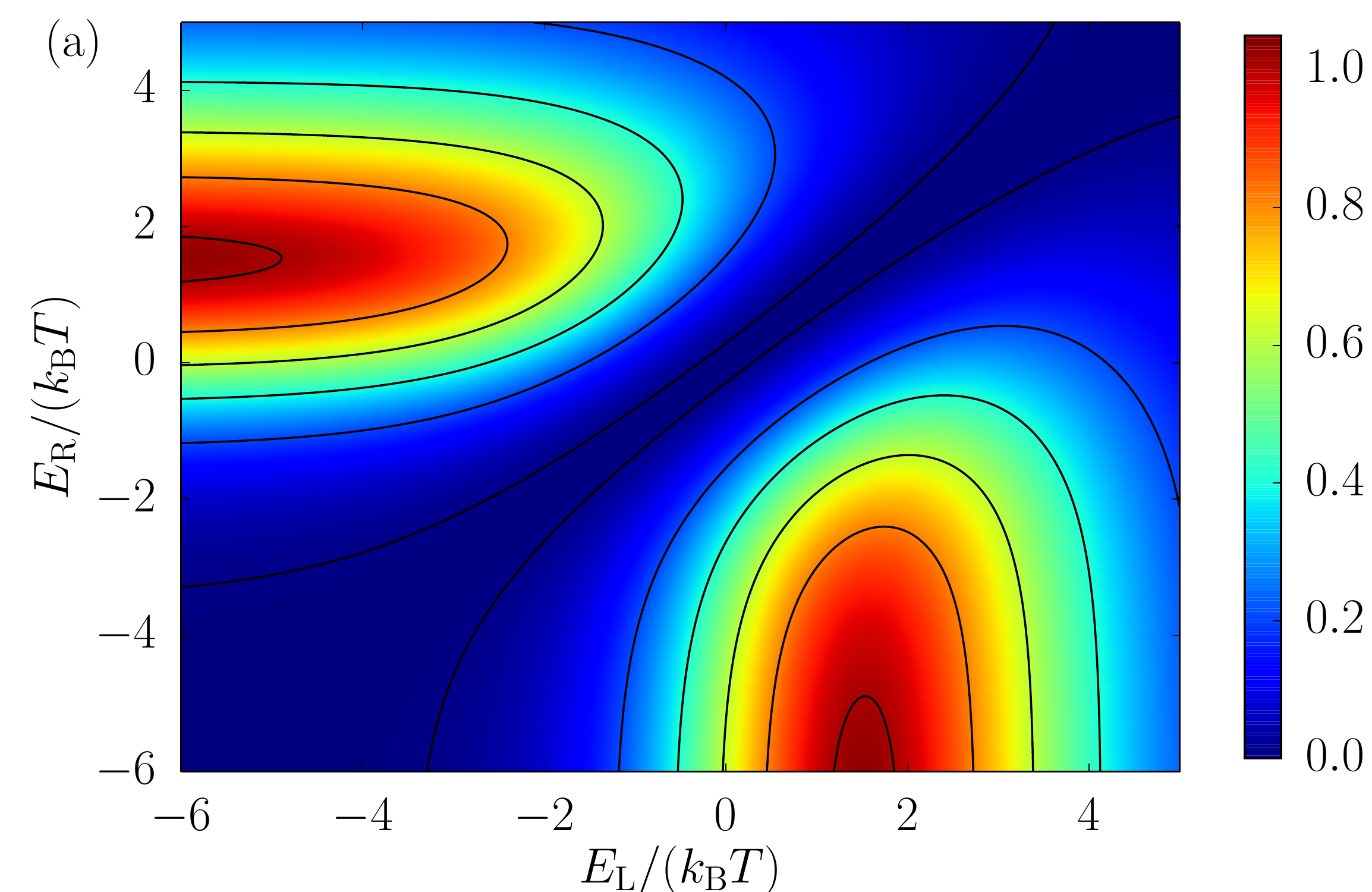}
	\includegraphics[width=.49\columnwidth]{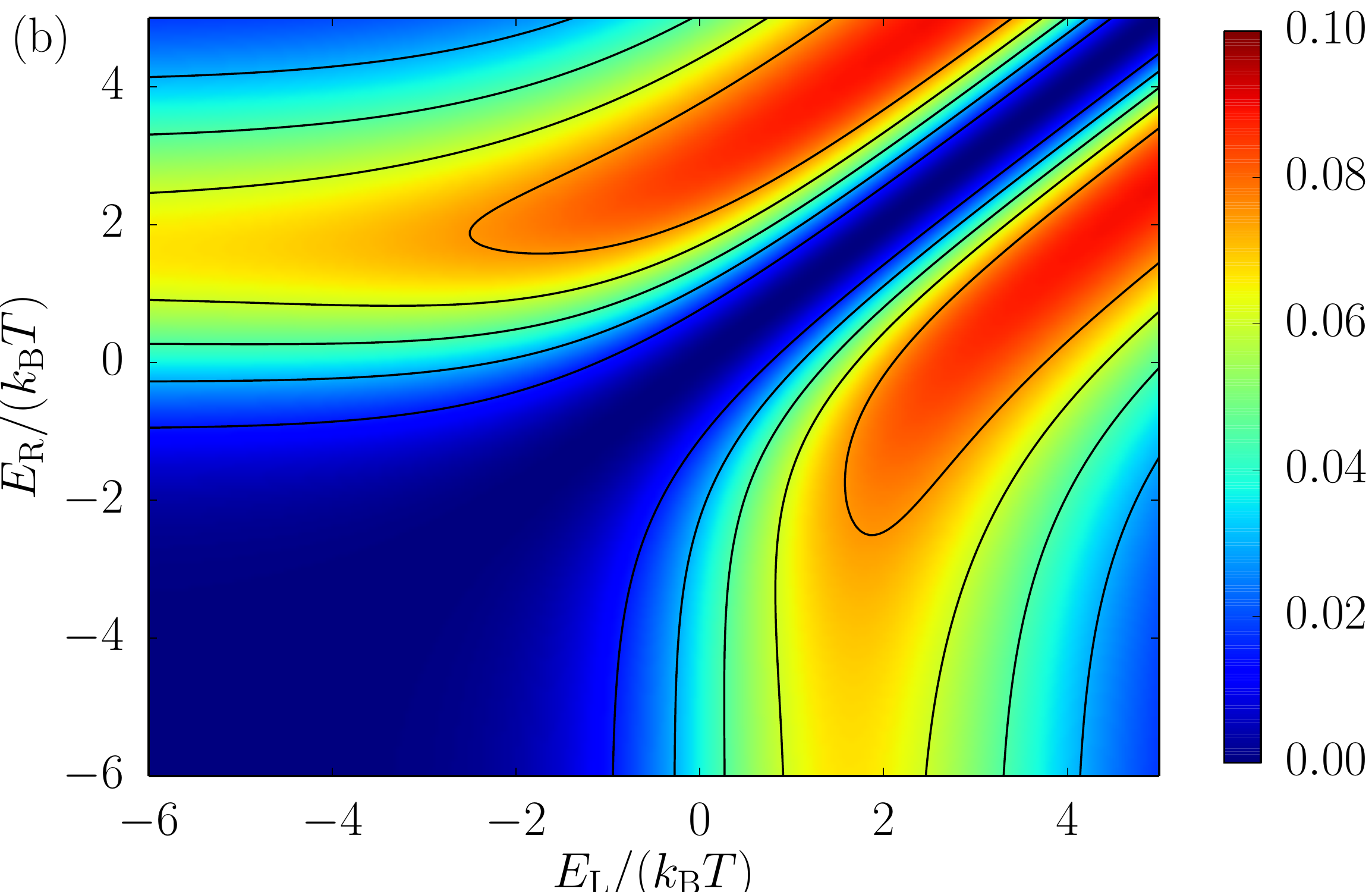}
	\includegraphics[width=.49\columnwidth]{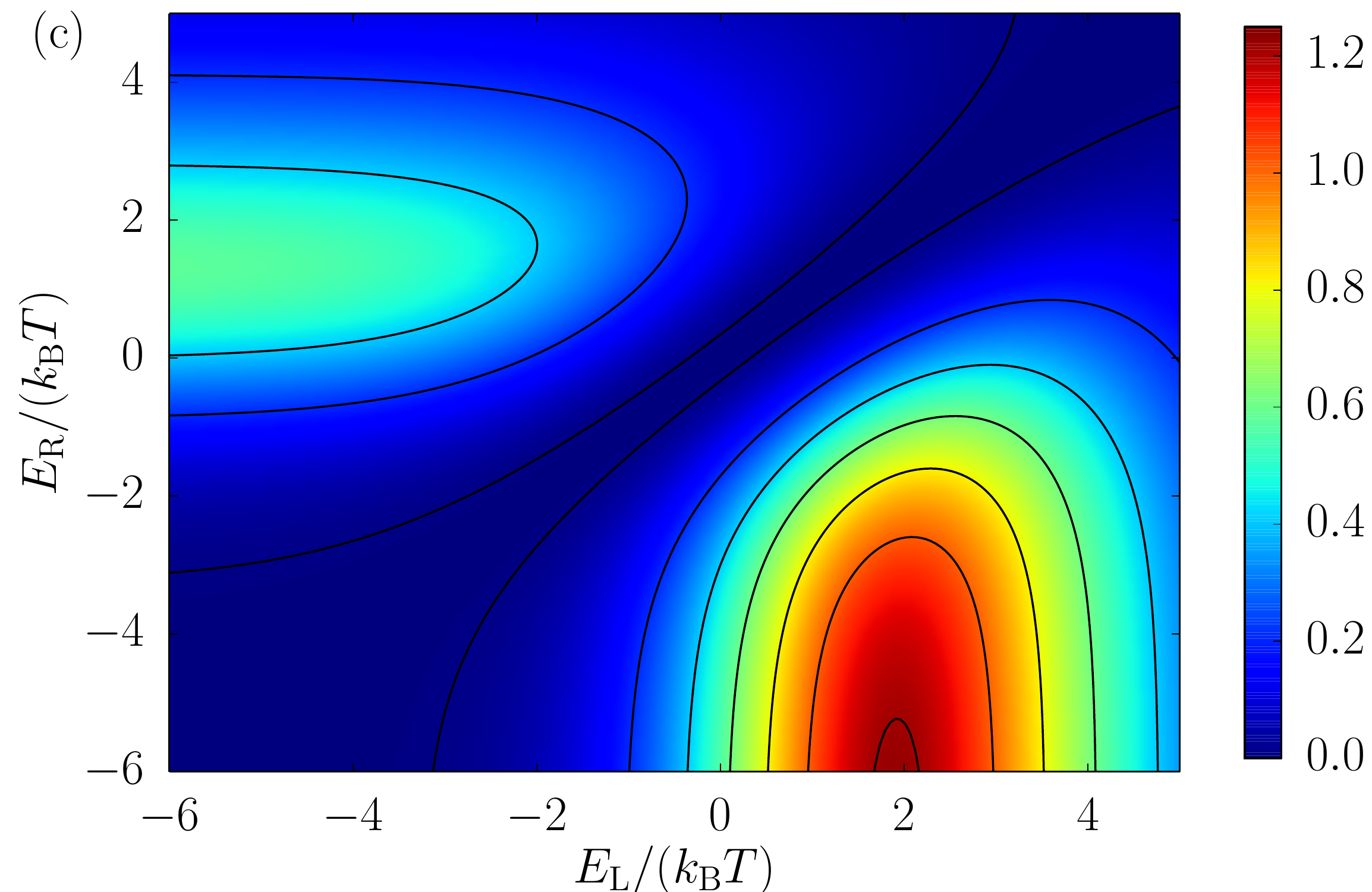}
	\includegraphics[width=.49\columnwidth]{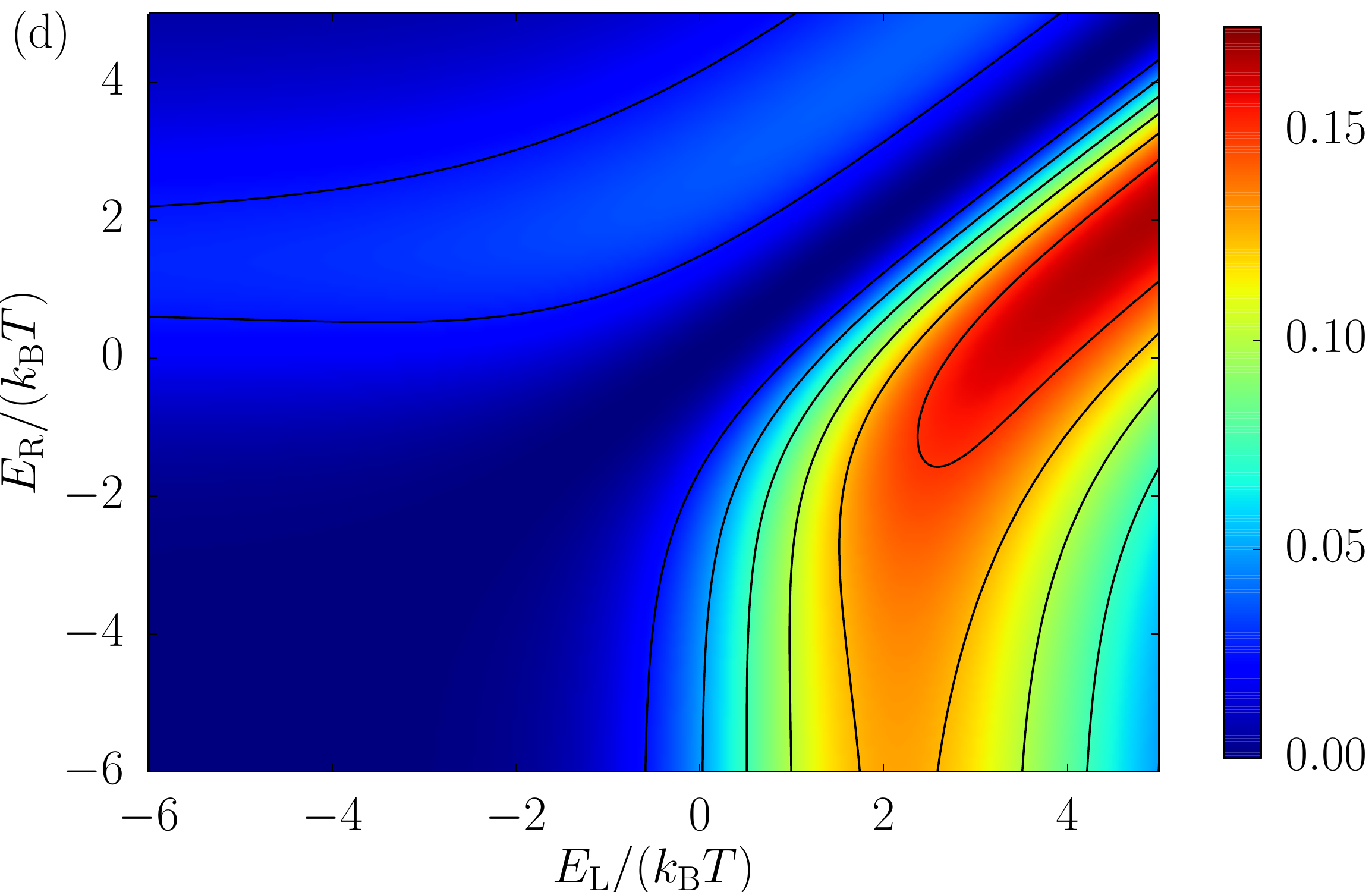}
	\caption{\label{fig:well_power}(a) Maximal output power in units of $\frac{\nu_2\mathcal A\Gamma}{2\hbar}(\frac{\kB \Delta T}{2})^2$ as a function of the threshold energies of the two quantum wells within linear response for a symmetric configuration $a=0$. (b) Efficiency at maximum power in units of the Carnot efficiency as a function of the threshold energies of the two quantum wells within linear response for a symmetric setup. (c) and (d) show the same as (a) and (b) but for an asymmetric device with $a=0.5$. (Reprinted with permission from~\cite{sothmann_powerful_2013}. Copyright 2013 IOP Publishing.)}
\end{figure}

In the following, we analyze the thermoelectric properties of the quantum-well heat engine in the linear response regime. For simplicity, we assume that both quantum wells are intrinsically symmetric, i.e. we have $\Gamma_{r1}=\Gamma_{r2}$. We parametrize the tunnel couplings as $\Gamma_{\text{L}1}=\Gamma_{\text{L}2}=(1+a)\Gamma$ and $\Gamma_{\text{R}1}=\Gamma_{\text{R}2}=(1-a)\Gamma$ where $\Gamma$ denotes the total coupling strength while $-1\leq a\leq1$ describes the asymmetry between the couplings of the left and right quantum well. We, furthermore, introduce the average temperature $T=(T_h+T_c)/2$ and the temperature difference $\Delta T=T_h-T_c$.

The charge current through the heat engine to linear order in the applied bias voltage $eV=\mu_\text{R}-\mu_\text{L}$ and in the applied temperature difference $\Delta T$ is given by 
\begin{equation}\label{eq:well_current}
	I_\text{L}=-I_\text{R}=\frac{e\nu_2\mathcal A\Gamma}{2\hbar}g_1\left(\frac{E_\text{L}}{\kB T},\frac{E_\text{R}}{\kB T}\right)\left[-eV-\kB\Delta T g_2\left(\frac{E_\text{L}}{\kB T},\frac{E_\text{R}}{\kB T}\right)\right],
\end{equation}
where we introduced the auxiliary functions
\begin{equation}
	g_1(x,y)=\frac{1-a^2}{2+(1-a)e^x+(1+a)e^y}
\end{equation}
and
\begin{equation}
	g_2(x,y)=x-y+(1+e^x)\log(1+e^{-x})-(1+e^y)\log(1+e^{-y}).
\end{equation}
From \eref{eq:well_current} we directly infer that a finite temperature bias $\Delta T$ can drive a charge current in the absence of an applied bias voltage $eV$. The direction of the charge current can be tuned by adjusting the threshold energies $E_\text{L}$ and $E_\text{R}$.

By applying a bias voltage $eV$ against the heat-driven current, the heat engine can perform work. The resulting output power $P=I_\text{L}V$ becomes maximal at half the stopping voltage where it takes the value
\begin{equation}
	P_\text{max}=\frac{\nu_2\mathcal A\Gamma}{2\hbar}\left(\frac{\kB\Delta T}{2}\right)^2 g_1\left(\frac{E_\text{L}}{\kB T},\frac{E_\text{R}}{\kB T}\right)g_2^2\left(\frac{E_\text{L}}{\kB T},\frac{E_\text{R}}{\kB T}\right).
\end{equation}
In order to evaluate the efficiency of heat to work conversion, again given by the ratio between output power and input heat, we need to calculate the heat current injected from the hot bath. At half the stopping voltage, it takes the form
\begin{equation}\label{eq:J}
	J=\frac{\nu_2\mathcal A\Gamma}{2\hbar}(\kB T)^2 \frac{\Delta T}{T} g_3\left(\frac{E_\text{L}}{\kB T},\frac{E_\text{R}}{\kB T}\right),
\end{equation}
where the function $g_3(x,y)$ satisfies $0<g_3(x,y)<2\pi^2/3$. Its complete analytical expression can be found in  Ref.~\cite{sothmann_powerful_2013}. 
Hence, the efficiency at maximum power is given by
\begin{equation}
	\eta_\text{maxP}=\frac{\eta_\text{C}}{4}\frac{g_1\left(\frac{E_\text{L}}{\kB T},\frac{E_\text{R}}{\kB T}\right)g_2^2\left(\frac{E_\text{L}}{\kB T},\frac{E_\text{R}}{\kB T}\right)}{g_3\left(\frac{E_\text{L}}{\kB T},\frac{E_\text{R}}{\kB T}\right)}.
\end{equation}

In the following, we discuss the output power and efficiency in more detail, starting with a symmetric setup $a=0$. In this case, both power and efficiency are symmetric with respect to an interchange of $E_\text{L}$ and $E_\text{R}$, cf. \fref{fig:well_power}. The output power takes its maximal value $P_\text{max}\approx\frac{\nu_2\mathcal A\Gamma}{2\hbar}\left(\frac{\kB \Delta T}{2}\right)^2$ when one level is deep below the equilibrium chemical potential while the other is located at approximately $1.5\kBT$ above it. An explanation for this behaviour will be given below.
The efficiency at maximum power takes its maximal value $\eta_\text{maxP}\approx0.1\eta_C$ when both levels are above the equilibrium chemical potential and satisfy $E_\text{L}\approx E_\text{R}\pm2\kBT$. However, for these parameters there is only an exponentially suppressed  number of electrons that can contribute to transport such that the output power in this regime becomes vanishingly small. For level positions that optimize the output power, we find an efficiency at maximum power of about $\eta_\text{maxP}\approx 0.07\eta_C$. 
% \textcolor{blue}{Hence, the quantum-well heat engine is significantly less efficient than a quantum-dot based setup in the limit of narrow resonances $\Gamma\ll\kBT$. (DO WE WANT TO SAY THIS?)}
Hence, the quantum-well heat engine is not as efficient as a quantum-dot based setup in the limit of narrow resonances $\Gamma\ll\kBT$.
The reason for this lies in the different energy-filtering properties of quantum dots and wells. Quantum dots with narrow resonances transmit energies only at a single energy. Hence, they reach the tight-coupling limit where heat and charge current are proportional to each other. In this situation, the efficiency at maximum power is then given by $\eta_\text{maxP}=\eta_C/2$.
Quantum wells on the other hand transmit any electron with an energy larger than the threshold voltage as in this case the energy $E$ can be decomposed into a part associated with the motion in the plane of the well and perpendicular to it, $E=E_z+E_\perp$. Hence, high-energy electrons can be transmitted through the well if most of their energy is in the perpendicular degrees of freedom such that $E_z$ matches the resonance condition. As a result, quantum wells act as much less efficient energy filters.

Given the rather weak energy filtering properties of quantum wells, it is quite surprising that the efficiency at maximum power is only a factor of three smaller than for quantum dots with level widths of the order of $\kBT$ -- the configuration that gives the largest output power, as discussed in Sec.~\ref{ssec:QDResult}. To understand this feature, we consider the situation depicted in \fref{fig:well_model}. The right quantum well has a threshold energy slightly above the equilibrium chemical potential. As the number of electrons with energies much larger than $E_\text{R}$ is exponentially small, it acts as a good energy filter.
For the left well, the energy filtering relies on a different mechanism. Electrons with energy $E$ can enter the cavity only if the associated state in the left reservoir is occupied, $f_\text{L}(E)>0$ and, at the same time, the corresponding state in the cavity is empty, $f_\text{C}(E)<1$. This defines an energy window of about $\kBT$ in which electrons are transmitted through the well. This explains why both quantum-dot and quantum-well based heat engines have similar efficiencies.

We now turn to the asymmetric case $a\neq0$. In this situation, the output power and efficiency are no longer symmetric with respect to an interchange of $E_\text{L}$ and $E_\text{R}$. In fact, as can be seen in \fref{fig:well_power}c) and d), we find that power and efficiency are strongly suppressed if $E_\text{L}<0$ and $E_\text{R}>0$ for an asymmetry $a>0$ (the roles of $E_\text{L}$ and $E_\text{R}$ are interchanged for $a<0$). However, for $E_\text{L}>0$ and $E_\text{R}<0$ we find that the power can be enhanced by up to $20\%$ for an asymmetry of about $a\approx 0.5$ while the efficiency at maximum power is even nearly doubled compared to the symmetric case.

We finally estimate what output power can be expected for a realistic device. For a GaAs based structure with an effective electron mass $m_*=0.067 m_e$, level width of $\Gamma=\kBT$ and asymmetry $a=0.5$ operating at room-temperature $T=\unit[300]{K}$, we obtain a maximal output power of $P_\text{max}=\unit[0.18]{W cm^{-2}}$ for a temperature bias of $\Delta T=\unit[1]{K}$. Hence, the quantum-well heat engine is about a factor of two more powerful than the previously discussed quantum-dot heat engine. We remark that materials with higher effective mass can yield even larger output powers.
Similarly to the quantum-dot case, we also find that a quantum-well heat engine is robust with respect to fluctuations of the threshold energies. As can be seen in \fref{fig:well_power}c) the output power is hardly affected at all by fluctuations of the right threshold energy as long as $-E_\text{R}\gg \kBT$ is fulfilled. Fluctuations of the left threshold energy are more crucial, but even here we find that variations of as much as $\kBT$ reduce the output power by only 20\%.

\subsubsection{Nonlinear response}
\begin{figure}
	\includegraphics[width=.49\columnwidth]{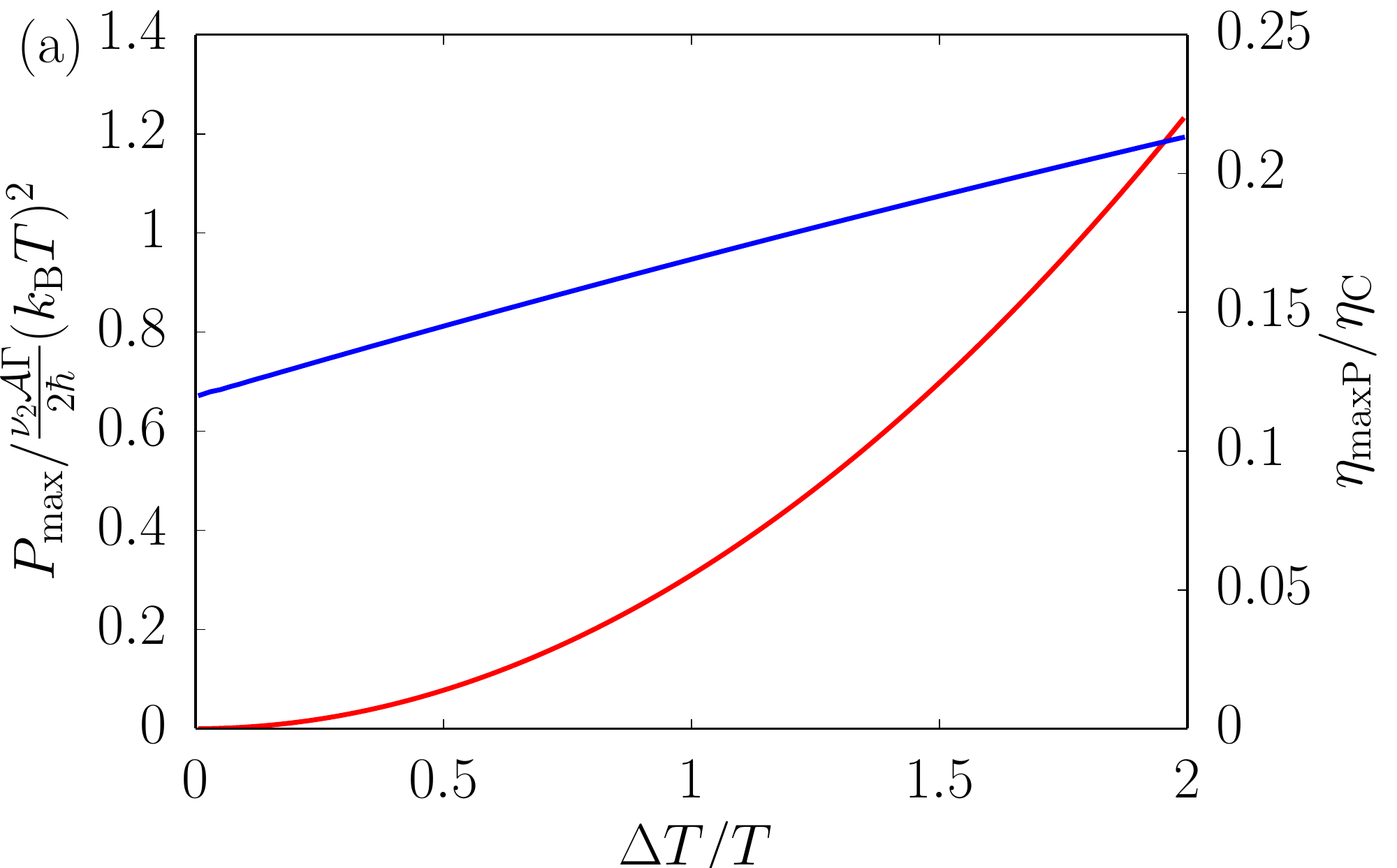}
	\includegraphics[width=.49\columnwidth]{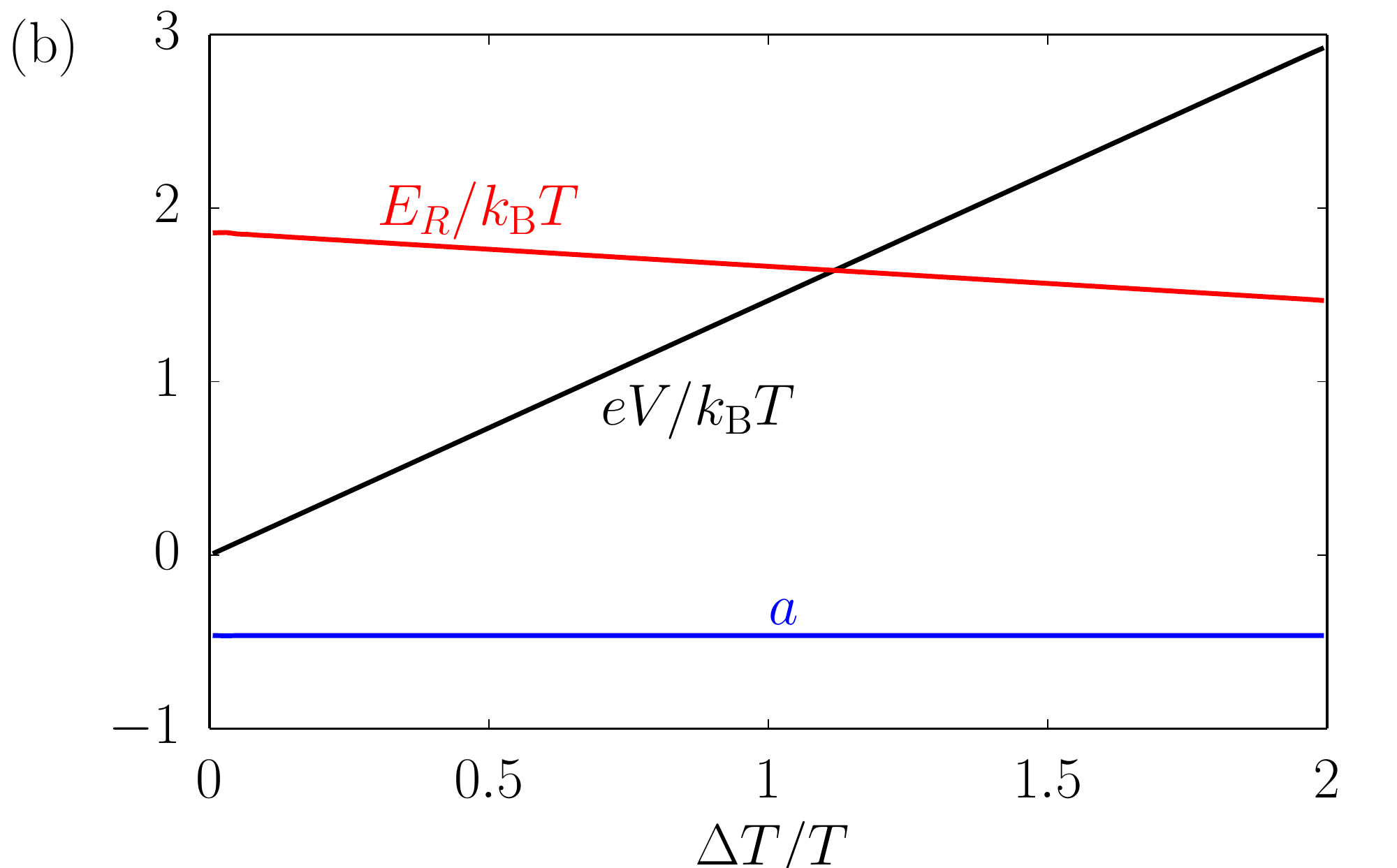}
	\caption{\label{fig:QW_nonlinear} (a) Maximized output power (red) and efficiency at maximum power (blue) as a function the temperature bias $\Delta T$. (b) Parameters that maximize the output power. (Reprinted with permission from~\cite{sothmann_powerful_2013}. Copyright 2013 IOP Publishing.)}
\end{figure}

We now turn to a discussion of the thermoelectric performance of our heat engine in the nonlinear regime. Similarly, to the quantum dot heat engine discussed in Sec.~\ref{ssec:QDResult}, we optimize the applied bias voltage, the threshold energies of the wells as well as the asymmetry $a$ for a given temperature bias $\Delta T$. The corresponding results are shown in \fref{fig:QW_nonlinear}b). The optimal bias voltage grows linearly with the temperature bias. The optimal asymmetry $a\approx -0.46$ is independent of $\Delta T$. The optimal threshold energy for the right quantum well decreases only slightly with the temperature bias while the threshold energy of the left well should be chosen as $-E_\text{L}\gg\kBT$ independent of $\Delta T$. The optimized power plotted in \fref{fig:QW_nonlinear}a) grows quadratically with the applied temperature difference and is approximately given by $P_\text{max}=0.3\frac{\nu_2\mathcal A\Gamma}{2\hbar}\left(\kB\Delta T\right)^2$. Due to the quadratic dependence on $\Delta T$ we obtain the same output power for a given value of $\Delta T$ in the linear and nonlinear regime. However, as the efficiency at maximum power grows linearly with the applied temperature bias, the device should be operated as much in the nonlinear regime as possible. In the extreme case of $\Delta T/T=2$, the quantum-well heat engine reaches an efficiency at maximum power of $\eta_\text{maxP}=0.22\eta_C$, i.e. it is about as efficient as the quantum-dot heat engine but delivers twice the power.

%------------------------------------------------------------------------------------------------------------------------------------------------------------------------
%#                                                                                                                                                                      #
%#                                                                                                                                                                      #
%#                                                                                                                                                                      #
%------------------------------------------------------------------------------------------------------------------------------------------------------------------------

\section{\label{sec:boson}Harvesting from bosonic sources}
So far, we discussed three-terminal heat engines based on electronic degrees of freedom only. In the following, we turn to a different class of energy harvesters where heat is injected from a third \emph{bosonic} reservoir. First, we will discuss several heat engines that are powered by phonons~\cite{entin-wohlman_three-terminal_2010,entin-wohlman_three-terminal_2012}. We will then analyze a magnon-driven heat engine that offers an additional spin degree of freedom, thereby establishing the connection to spin caloritronics~\cite{sothmann_magnon-driven_2012}. Finally, we will discuss heat engines driven by photons. These photons can either be absorbed  from the electromagnetic environment~\cite{ruokola_theory_2012} or be specifically emitted by the hot source into a resonant superconducting microwave cavity that serves as a quantum bus for heat exchanged between spatially separated mesoscopic conductors~\cite{bergenfeldt_hybrid_2014}.

%------------------------------------------------------------------------------------------------------------------------------------------------------------------------
%#                                                                                                                                                                      #
%#                                                                                                                                                                      #
%#                                                                                                                                                                      #
%------------------------------------------------------------------------------------------------------------------------------------------------------------------------

\subsection{\label{ssec:phonon}Phonon harvesting}
\begin{figure}
	\centering
	\includegraphics[width=.49\textwidth]{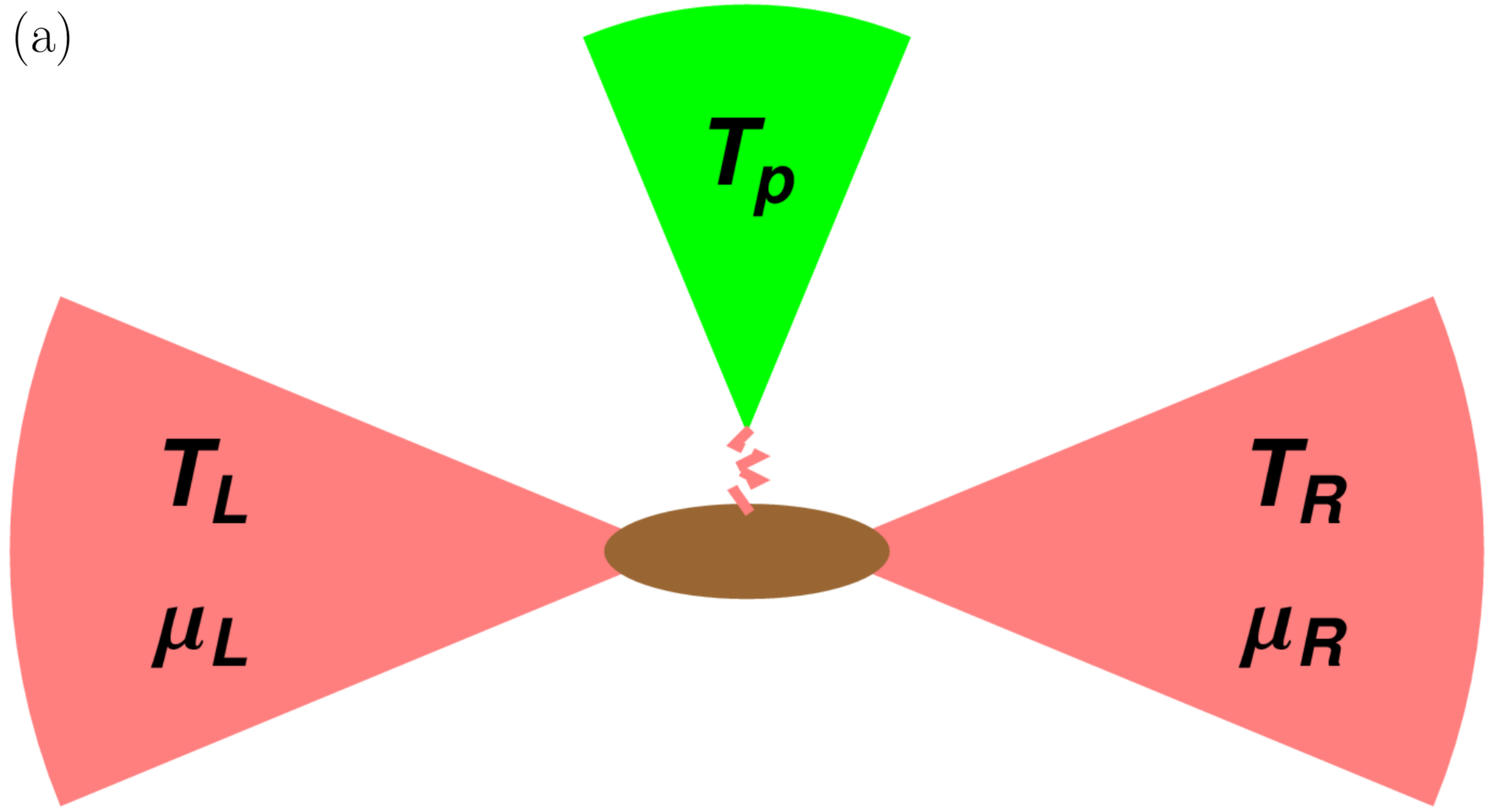}
	\includegraphics[width=.49\textwidth]{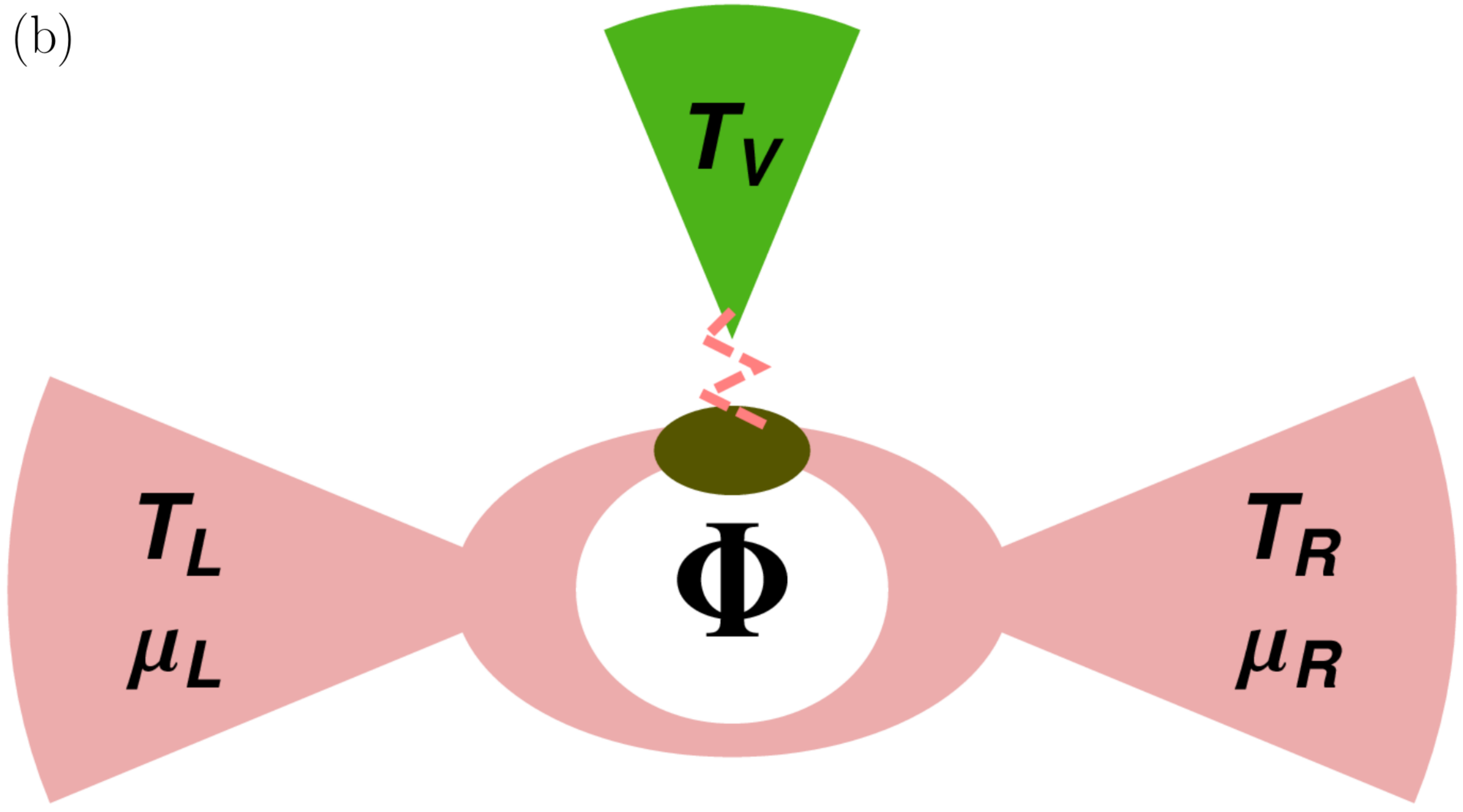}
	\caption{\label{fig:Phonon_model}(a) A quantum dot (brown) is coupled to two electronic reservoirs (pink) with temperatures $T_L$ and $T_R$ as well as to a phonon bath (green) with temperature $T_P$. (Reprinted with permission from~\cite{entin-wohlman_three-terminal_2010}. Copyright 2010 American Physical Society.) (b) Aharonov-Bohm ring with a quantum dot embedded in one of the arms. (Reprinted with permission from~\cite{entin-wohlman_three-terminal_2012}. Copyright 2012 American Physical Society.)}
\end{figure}

As a first example of a boson-driven heat engine, we consider a single-level quantum dot coupled to two electronic reservoirs as well as a to a local phonon mode, cf.~\fref{fig:Phonon_model}a)~\cite{entin-wohlman_three-terminal_2010}. Experimentally, such a system can be realized in a single-molecule junction where generically the electron localized on the molecule couples strongly to the molecular vibrations. This electron-phonon coupling gives rise to a number of interesting phenomena such as Franck-Condon blockade of transport at small bias voltages~\cite{koch_franck-condon_2005,koch_theory_2006,leturcq_franck-condon_2009}, nonequilibrium phonon distributions~\cite{koch_current-induced_2006,cavaliere_phonon_2008}, lasing of optical phonons~\cite{okuyama_lasing_2013}, two-terminal thermoelectric effects~\cite{koch_thermopower_2004,leijnse_nonlinear_2010} and allows the study of fluctuation theorems in the presence of fermionic and bosonic degrees of freedom~\cite{krause_incomplete_2011,simine_vibrational_2012,schaller_single-electron_2013}. Here, we demonstrate that such a setup can also serve as a heat engine driven by the temperature bias between the electronic and phononic reservoirs~\cite{entin-wohlman_three-terminal_2010}. The Hamiltonian of the system can be written as the sum of three parts, 
\begin{equation}
	H=\sum_r H_r+H_\text{dot}+H_\text{tun}.
\end{equation}
The first term describes the left, $r=L$, and right, $r=R$, electronic reservoir in terms of noninteracting, spinless electrons (we neglect the spin degree of freedom in the following as it only leads to a trivial renormalization of the tunnel couplings defined below)
\begin{equation}
	H_r=\sum_{\vec k}\varepsilon_{r\vec k}a_{r\vec k}^\dagger a_{r\vec k}.
\end{equation}
The two electronic reservoirs are at temperature $T_r$ and chemical potential $\mu_r$, respectively. For later convenience, we parametrize the temperatures and chemical potentials as $T_{L/R}=T\pm\Delta T/2$ and $\mu_{L/R}=\mu\pm\delta \mu/2$.

The molecule inside the junction is described by the Hamiltonian
\begin{equation}
	H_\text{dot}=\varepsilon_0 c^\dagger c+\omega_0\left(b^\dagger b+\frac{1}{2}\right)+\gamma(b+b^\dagger) c^\dagger c.
\end{equation}
The molecule hosts a single level relevant for transport with energy $\varepsilon_0$. It is coupled to a single dispersionless vibrational mode with energy $\omega_0$ via coupling strength $\gamma$. While in principle the phonon distribution could be a nonequilibrium one, in the following we consider the situation where it is thermal with temperature $T_P=T+\Delta T_P$ due to the coupling to the environment.
Finally, tunneling between the dot and the electrodes is given by
\begin{equation}
	H_\text{tun}=\sum_{r\vec k}t_r a_{r\vec k}^\dagger c+\text{H.c.}
\end{equation}
where the tunnel matrix elements $t_r$ are related to the tunnel coupling strength as $\Gamma_r(\omega)=2\pi\sum_{\vec k}|t_r|^2\delta(\omega-\varepsilon_{r\vec k})$.

For the system at hands, the charge current $I$ through the molecule as well as the electronic and phononic heat currents, $J^\text{el}$ and $J^\text{ph}$, can be evaluated using a nonequilibrium Green's function approach that takes into account the tunneling between the dot and the leads exactly and performs a perturbative expansion in the electron-phonon coupling $\gamma$ up to second order. In linear response, the currents are related to the corresponding thermodynamic forces $\delta\mu/e$, $\Delta T/T$ and $\Delta T_P/T$ via the Onsager matrix 
\begin{equation}
	\vec M
	=
	\left(
	\begin{array}{ccc}
		G & K & X^p \\
		K & K_2 & \tilde X^p \\
		X^p & \tilde X^p & C^p
	\end{array}
	\right).
\end{equation}
The full analytic expressions for the Onsager coefficients can be found in Ref.~\cite{entin-wohlman_three-terminal_2010}. While the coefficients $G$, $K$ and $K_2$ contain contributions that are both due to elastic and inelastic transport through the molecule, the other Onsager coefficients result from inelastic processes only. 
The coefficient $X^p$ describes a charge current response to a temperature difference between phonons and electrons. In order to have a finite $X^p$, both the left-right as well as the particle hole symmetry need to be broken, i.e. $\Gamma_r(\omega)$ must be energy-dependent and fulfill $\Gamma_\text{L}(\omega)\neq \Gamma_\text{R}(\omega)$. Similarly, the electronic heat current due to a temperature difference between electrons and phonons, characterized by the Onsager coefficient $\tilde X^p$, requires a breaking of left-right and particle-hole symmetry.

An extension of the setup we just discussed consists of an Aharonov-Bohm geometry with a molecular quantum dot embedded in one of the arms~\cite{entin-wohlman_three-terminal_2012}, cf.~\fref{fig:Phonon_model}b).  Transport through the reference arm of the interferometer is described by the new contribution to the Hamiltonian
\begin{equation}
	H_\text{LR}=\sum_{\vec k\vec k'}t_\text{LR}e^{i\phi}a_{\text{L}\vec k}^\dagger a_{\text{R}\vec k'}+\text{H.c.},
\end{equation}
where $\phi$ denotes the Aharonov-Bohm flux through the ring. Due to the Aharonov-Bohm flux, the Onsager matrix becomes $\phi$-dependent, $\vec M(\phi)=\vec M^T(-\phi)$, such that the Onsager relations~\cite{onsager_reciprocal_1931,casimir_onsagers_1945,jacquod_onsager_2012} are satisfied. The analytic expressions for the Onsager coefficients that are given in Ref.~\cite{entin-wohlman_three-terminal_2012} exhibit three different types of flux dependence. First, there are contributions proportional to $\cos\phi$ that arise from interference contributions. Second, there are terms proportional to $\cos2\phi$ due to contributions from time-reversed paths. Third, there are $\sin\phi$ contributions that emerge from the coupling to the phonons. While the former two contributions are even in the flux, the latter one is odd. The odd contributions in the off-diagonal Onsager coefficients can potentially lead to a large efficiency at maximum power~\cite{benenti_thermodynamic_2011,brandner_strong_2013,balachandran_efficiency_2013,brandner_multi-terminal_2013,stark_classical_2014,sothmann_quantum_2014,sanchez_chiral_2014} that overcomes the linear response limit $\eta_\text{maxP}=\eta_C/2$ that exists for time-reversal symmetric systems~\cite{van_den_broeck_thermodynamic_2005}. However, for the system at hands, these efficiency bounds have not yet been investigated.

The idea of phonon-driven thermoelectric transport in multi-terminal devices was further discussed in the context of phonon-assisted hopping~\cite{jiang_thermoelectric_2012}. Furthermore, a p-i-n-diode structure  that drives a charge current by harvesting energy from a hot phonon source was discussed in Ref.~\cite{jiang_three-terminal_2013}. It was theoretically estimated that such a device based on a Bi$_2$Te$_3$/Si superlattice can have a figure of merit larger than 1 when operated at room temperature, thus making it a promising candidate for energy harvesting applications.

%------------------------------------------------------------------------------------------------------------------------------------------------------------------------
%#                                                                                                                                                                      #
%#                                                                                                                                                                      #
%#                                                                                                                                                                      #
%------------------------------------------------------------------------------------------------------------------------------------------------------------------------

\subsection{\label{ssec:magnon}Magnon harvesting}
\begin{figure}
	\centering\includegraphics[width=.5\columnwidth]{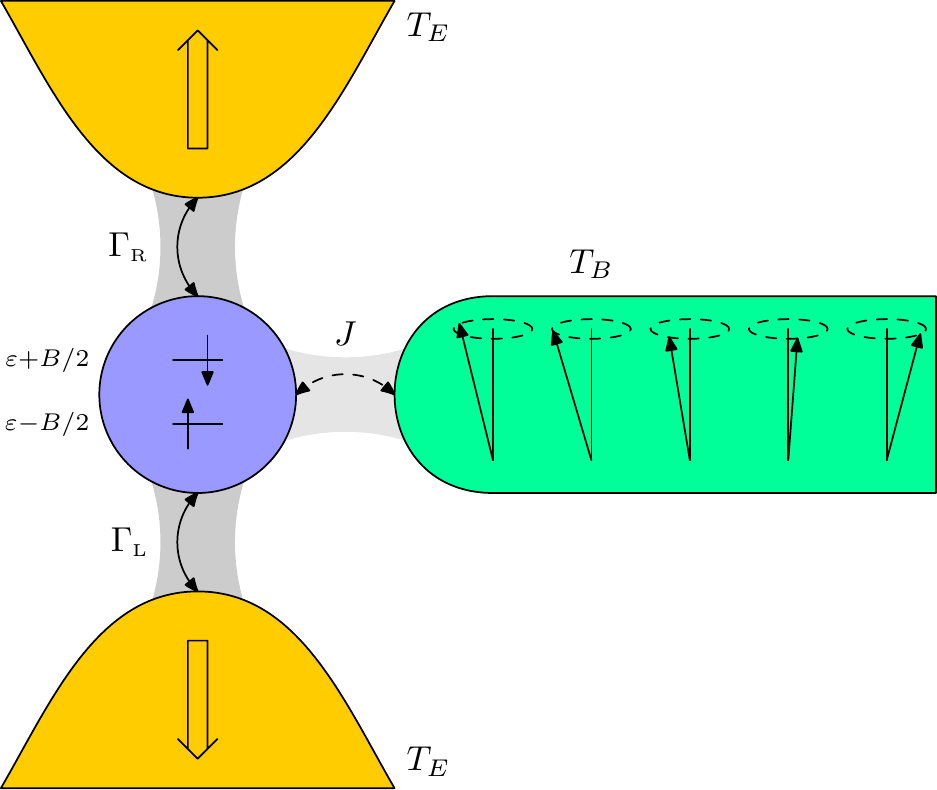}
	\caption{\label{fig:Magnon_Model}Schematic sketch of a magnon-driven quantum-dot heat engine. A single-level quantum dot (blue) is tunnel-coupled to two ferromagnetic metals (yellow). In addition, it is also exchange-coupled to a ferromagnetic insulator (green) serving as a source of spin waves. (Reprinted with permission from~\cite{sothmann_magnon-driven_2012}. Copyright 2012 IOP Publishing.)}
\end{figure}

We now turn to a quantum-dot heat engine that is driven by spin waves from a ferromagnetic insulator~\cite{sothmann_magnon-driven_2012}, cf. \fref{fig:Magnon_Model}. Compared to the phonon-driven setups that we discussed in the previous section, the magnon-driven heat engine offers a number of advantages. First of all, magnons are potentially easier to control than phonons. As phonons exist in any material, it is hard to avoid leakage heat currents from the hot phonon bath to the cold electronic reservoirs. Magnons, in contrast, exist only in magnetic materials and couple via short-range exchange interactions which facilitates coupling them only to the quantum dot degrees of freedom. Another advantage of the magnon-heat engine is that it does not rely on energy-dependent tunnel couplings. Instead, here the necessary asymmetry between electrons and holes is introduced into the system by the spin-dependence of the tunnel barriers. Finally, the magnon-driven heat engine provides an example of a spin caloritronic heat engine~\cite{bauer_spin_2012} that allows to drive spin-polarized charge currents as well as pure spin currents by thermal gradients. Alternative spintronic heat engines based on nanowires with domain walls have been proposed in Ref.~\cite{bauer_nanoscale_2010,kovalev_thermomagnonic_2012}.

A related setup has been discussed in Ref.~\cite{sothmann_influence_2010} where a system consisting of a quantum dot coupled to ferromagnetic electrodes with spin waves has been taken into account. Keeping the magnons inside the ferromagnets at a different temperature than the electrons, e.g., by microwave excitation of spin precession, gives rise to spin-polarized charge currents. While this effect was termed magnon-assisted transport in Ref.~\cite{sothmann_influence_2010} in analogy to phonon-assisted tunneling, it can also be viewed as a three-terminal thermoelectric device.

We consider a single-level quantum dot in the Coulomb-blockade regime tunnel coupled to two ferromagnetic electrodes $r=\text{L,R}$ at temperature $T_E$ as well as exchange coupled to a ferromagnetic insulator that serves as a source of magnons with temperature $T_B$.

The ferromagnetic electrodes are modeled in the spirit of a Stoner model as a noninteracting electron gas with a constant but spin-dependent density of states $\rho_{r\sigma}$. It is related to the spin polarization $p_r=(\rho_{r+}-\rho_{r-})/(\rho_{r+}+\rho_{r-})$ that varies between $p_r=0$ for a normal metal and $p_r=1$ for a half-metallic ferromagnet. In the following, we assume identical polarizations for the two leads, $p_r=p$.

The ferromagnetic insulator is modeled as a Heisenberg chain of exchange-coupled spins. Using a Holstein-Primakoff transformation, it can be described in terms of bosonic operators that create and annihilate magnons. At low temperatures where the average magnon number is small, the ferromagnetic insulator behaves as a noninteracting magnon gas with dispersion relation $\omega_{\vec q}$.

The quantum dot has a single spin-split level with energy $\varepsilon_\sigma=\varepsilon\pm B/2$. The Zeeman splitting $B$ due to an externally applied magnetic field determines the energy of the magnons that the heat engine can harvest. The Coulomb energy $U$ that is required to occupy the dot with two electrons at the same time is assumed to be infinite. We remark that taking into account a finite value of $U$ does not give rise to qualitatively different results.

Tunneling between the dot and the ferromagnetic electrodes is characterized by the spin-dependent tunnel couplings $\Gamma_{r\sigma}$. For later convenience we also introduce the total tunnel coupling strength $\Gamma_r=\Gamma_{r\up}+\Gamma_{r\down}$. The coupling between the dot and the ferromagnetic insulator is due to an exchange interaction with spectral weight $J(\omega)$ that flips the spin of the dot and emits or absorbs a magnon in the insulator. As in the following we only need the spectral weight evaluated at the Zeeman splitting of the dot, we will omit its energy dependence and write $J(\omega)=J$.

Transport through the system is described via a standard master equation approach that integrates out the noninteracting fermionic and bosonic degrees of freedom in the reservoirs and characterizes the quantum dot by its reduced density matrix. Within this approach, the spin-resolved electron and magnon currents can then be evaluated in a straightforward way. We count particle currents as positive when they flow from the reservoir into the dot.

\begin{figure}
	\centering\includegraphics[width=.475\columnwidth]{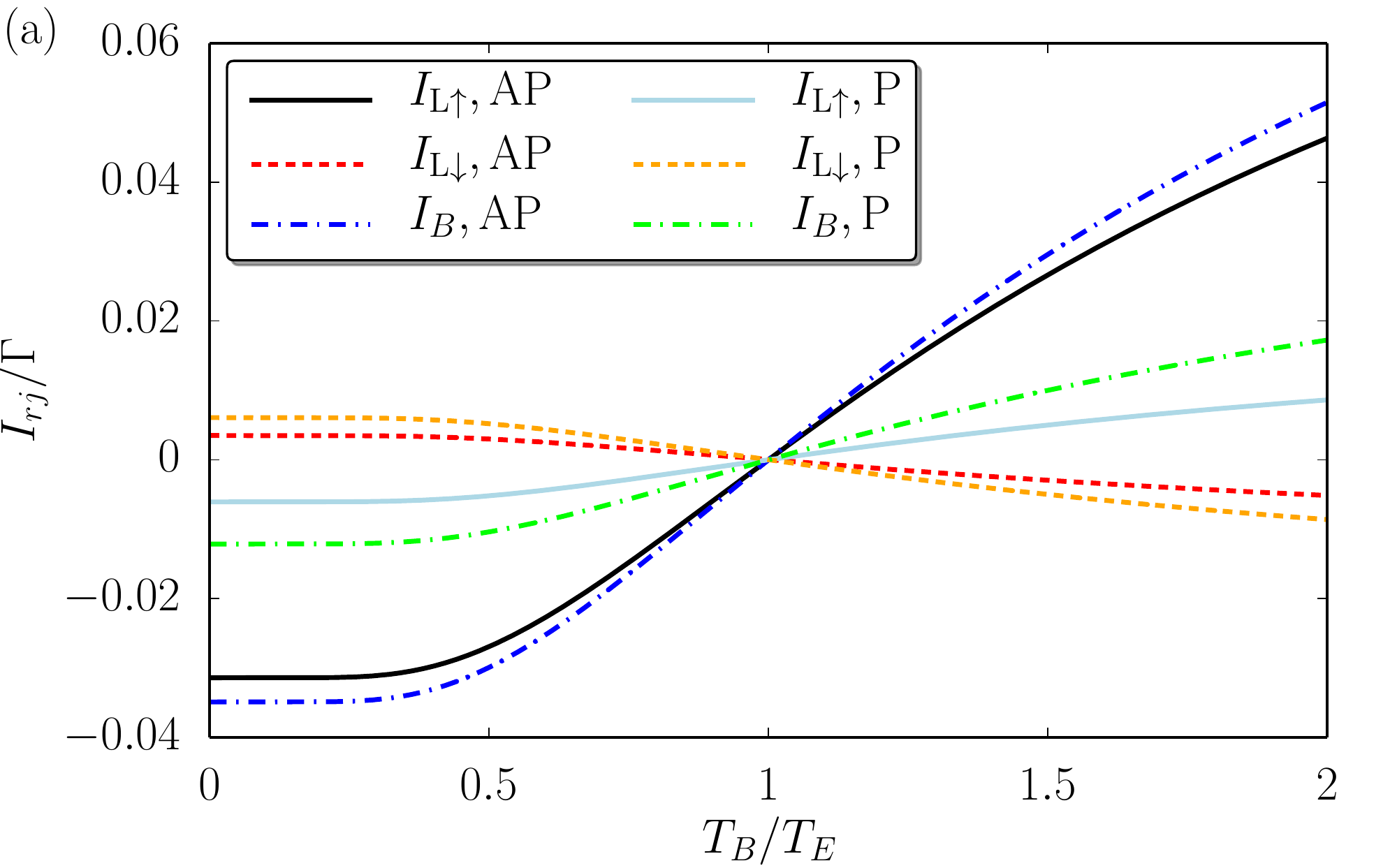}
	\includegraphics[width=.475\columnwidth]{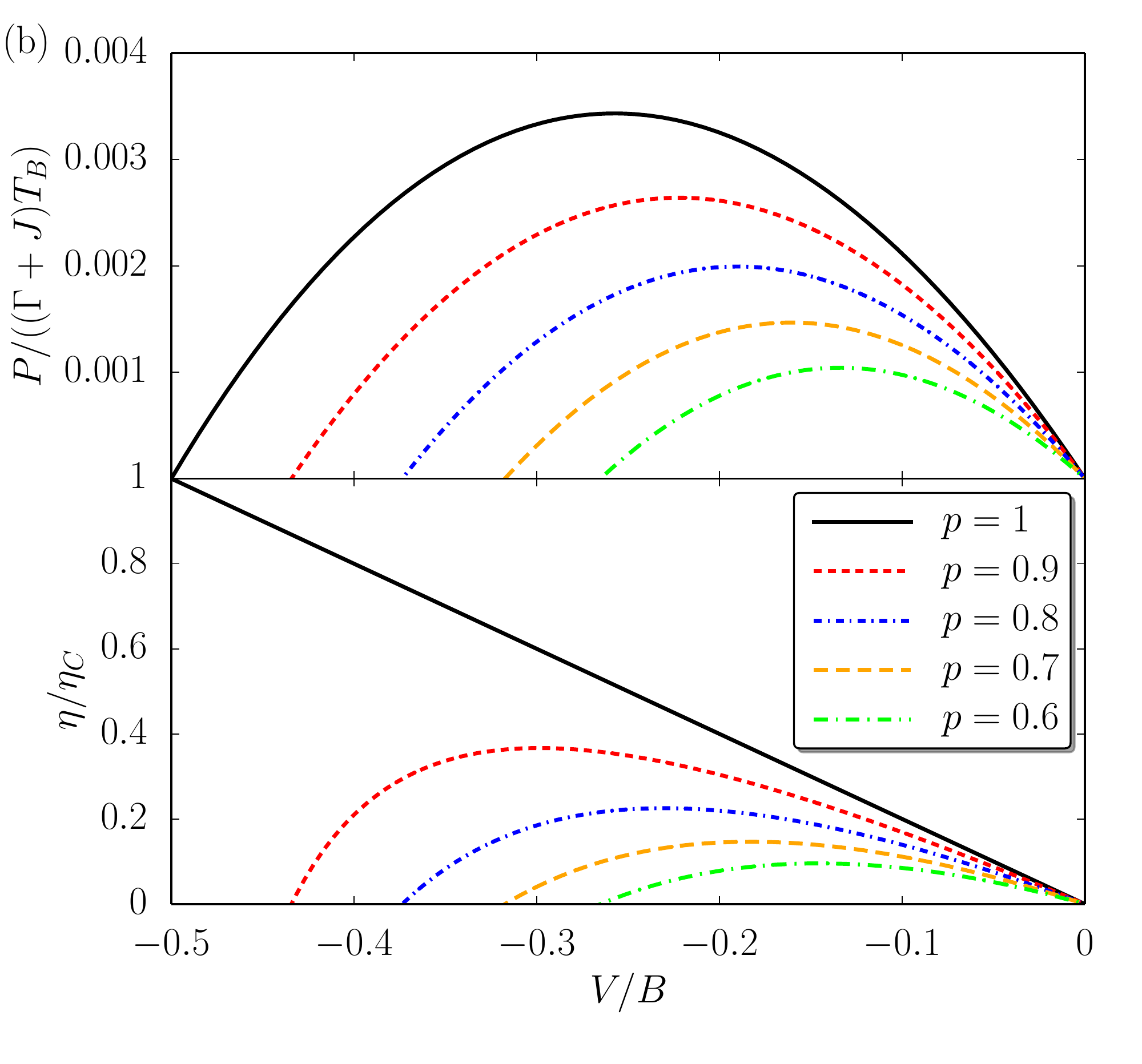}
	\caption{\label{fig:Magnon_Current}(a) Spin-resolved electron and magnon currents as a function of the magnon temperature for parallel and antiparallel geometry. Parameters are $J=\Gamma$, $B=2T_E$, $\varepsilon=0$ and $p=0.8$. (b) Power and efficiency as a function of applied bias voltage for different polarizations and other parameters as in the left panel. (Reprinted with permission from~\cite{sothmann_magnon-driven_2012}. Copyright 2012 IOP Publishing.)}
\end{figure}
We now turn to a discussion of the thermoelectric performance of the magnon-driven heat engine. \Fref{fig:Magnon_Current} shows the spin-resolved electron and magnon currents that flow in response to a temperature difference between electrons and magnons. For parallely magnetized electrodes, we find that as soon as $T_B\neq T_E$, we have finite electron and magnon currents flowing through the dot. For any temperature bias, we observe that the currents of spin up and spin down electrons have equal magnitude but opposite sign, $I_{\text{L}\up}=-I_{\text{L}\down}$. Hence, the total charge current $I_c=I_{\text{L}\up}+I_{\text{L}\down}$ vanishes while the spin current $I_s=I_{\text{L}\up}-I_{\text{L}\down}$ is finite, i.e. a pure spin current is generated. The physical picture behind this is that spin up electrons tunnel into the dot, absorb a magnon to flip their spin and then tunnel out again. On average, for each spin up electron entering the dot through a given tunnel barrier, there is a spin down electron leaving the dot through the same barrier such that $I_c=0$ and $I_s\neq0$. For antiparallel magnetizations of the ferromagnetic electrodes, the magnitudes of $I_{\text{L}\up}$ and $I_{\text{L}\down}$ are no longer equal to each other such that now a finite, spin-polarized charge current flows through the dot. It arises as spin up electrons preferably tunnel in from the left electrode whereas spin down electrons preferably tunnel out to the right electrode due to the spin-dependent tunnel couplings.

In the following, we focus on the antiparallel case and discuss the output power $P$ and efficiency $\eta$ of heat to work conversion when an external bias voltage $V$ is applied against the thermally driven current. \Fref{fig:Magnon_Current} shows the output power as a function of $V$ for different polarizations. It grows from $P=0$ at vanishing bias to a maximal value and then drops down again to zero at the stopping voltage. Furthermore, it grows as the polarization is increased because the energy filtering properties of the heat engine become more pronounced, reaching the tight-coupling limit for $p=1$. The efficiency $\eta$ shown in \fref{fig:Magnon_Current} exhibits qualitatively different behaviour for $p\neq 1$ and $p=1$. In the former case, it grows from $\eta=0$ at $V=0$ to a finite value and goes down to zero at the stopping voltage. We remark that the maximal efficiency in general occurs for a different bias voltage than the maximal power. For $p=1$, the efficiency just grows linearly with the applied voltage and reaches the Carnot efficiency at the stopping voltage as expected in the tight-coupling limit. The corresponding efficiency at maximum power $\eta_\text{maxP}$ is given by $\eta_C/2$ in the linear response regime while it satisfies $\eta_C/2\leq\eta_\text{maxP}\leq \eta_C/(2-\eta_C)$ in the nonlinear regime.

%------------------------------------------------------------------------------------------------------------------------------------------------------------------------
%#                                                                                                                                                                      #
%#                                                                                                                                                                      #
%#                                                                                                                                                                      #
%------------------------------------------------------------------------------------------------------------------------------------------------------------------------

\subsection{\label{ssec:photon}Photon harvesting}
Apart from phonons and magnons, one can also make use of photons to drive mesoscopic heat engines. Heat transfer due to photons in nanoscale circuits has been studied theoretically~\cite{schmidt_photon-mediated_2004,pekola_normal-metal-superconductor_2007,peltonen_brownian_2011,ojanen_photon_2007,ojanen_mesoscopic_2008,ruokola_thermal_2009,pascal_circuit_2011} as well as been experimentally observed~\cite{meschke_single-mode_2006,timofeev_electronic_2009}. Here, we present two different examples of photon harvesting. In the first case, a quantum-dot based setup is used to harvest microwave photons from the electromagnetic environment~\cite{ruokola_theory_2012}. A similar proposal used to harvest energy from visible light was analyzed in Ref.~\cite{rutten_reaching_2009}. In the second example, we consider a system of two double quantum dots connected via a superconducting microwave cavity. The latter serves as a quantum bus for heat flow between the quantum dots. Both types of setup offer the advantage of spatially separating the hot and the cold part of the heat engine. This potentially reduces leakage heat currents due to substrate phonons and can therefore help to achieve highly efficient heat engines.

\subsubsection{Photons from electromagnetic environment}
\begin{figure}
	\centering\includegraphics[width=.5\textwidth]{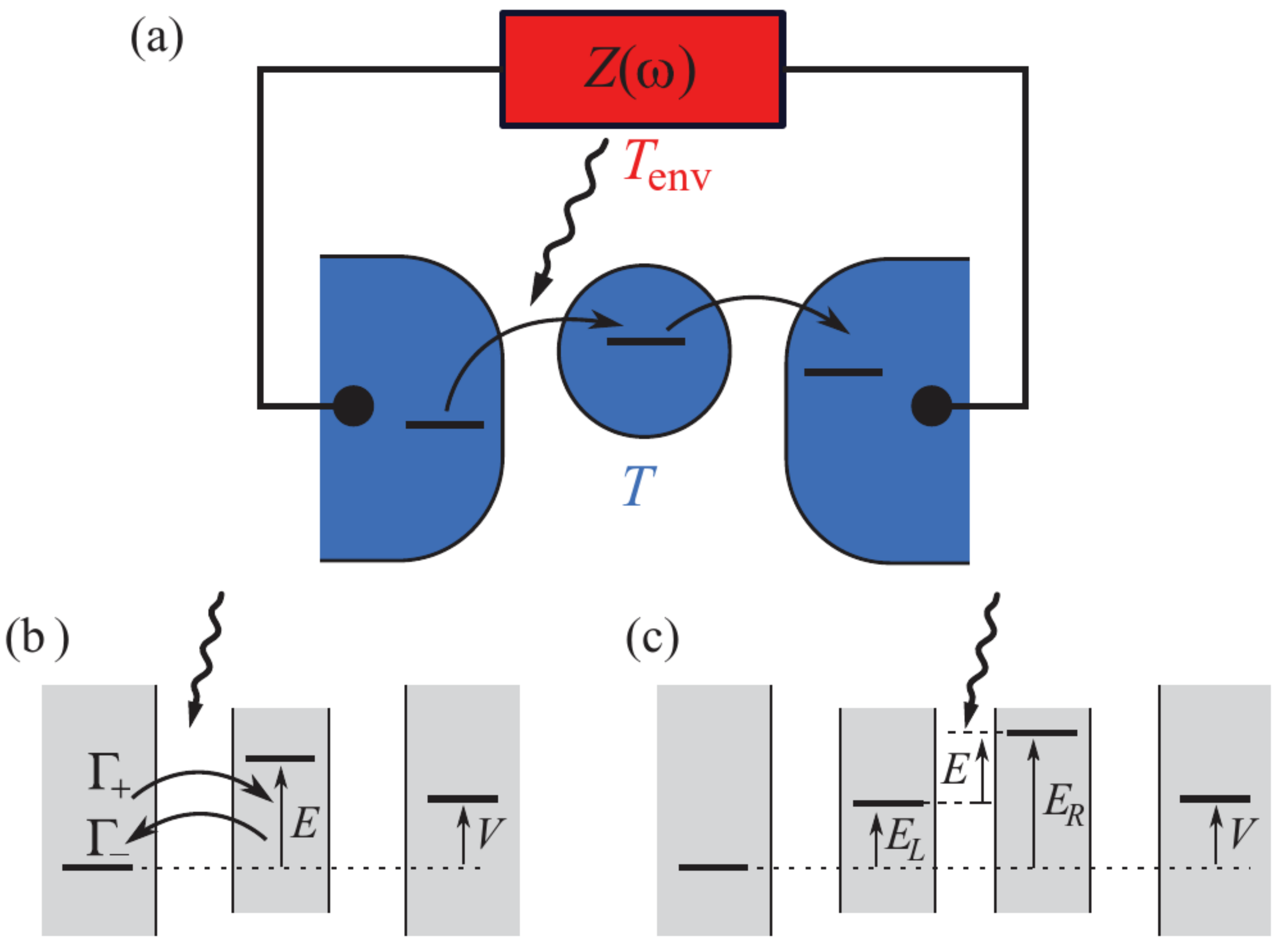}
	\caption{\label{fig:EM_Harvesting}(a) Heat engine that harvests energy from the electromagnetic environment. A quantum dot structure is coupled to two electronic reservoirs at temperature $T$. The circuit connecting the two leads contains an impedance $Z(\omega)$ at the environment temperature $T_\text{env}$. Realizations of the heat engine with a single dot and a double dot are shown in (b) and (c), respectively. (Reprinted with permission from~\cite{ruokola_theory_2012}. Copyright 2012 American Physical Society.)}
\end{figure}

We consider a mesoscopic heat engine as discussed in Ref.~\cite{ruokola_theory_2012} that consists of one or two quantum dots coupled to electronic reservoirs at temperature $T$. The electronic reservoirs are connected to each other via an external circuit with impedance $Z(\omega)$ that is kept at temperature $T_\text{env}$. We assume that the tunnel coupling between the dots and the leads is weak, such that a lowest order description of tunneling is valid. We, furthermore, assume that the relaxation time of the electromagnetic environment is much shorter than the average time between subsequent tunneling events. In this situation, transport through the system can be described by $P(E)$ theory~\cite{ingold_charge_1992}. The rate for an electron transition from system $i$ to $j$ is then given by 
\begin{equation}\label{eq:P(E)transition}
	\Gamma_{i\to j}=2\pi|t|^2\int d\varepsilon_i d\varepsilon_j \rho_i(\varepsilon_i-\mu_i) \bar \rho_j(\varepsilon_j-\mu_j) P(\varepsilon_i-\varepsilon_j).
\end{equation}
Here, $t$ denotes the tunnel matrix element for a transition from $i$ to $j$. The density of states for electrons $\rho(\varepsilon)$ is given by $\delta(\varepsilon)$ for a quantum dot with discrete levels and by $\nu_i f(\varepsilon)$ for a metallic island or an electrode where $\nu_i$ denotes the density of states at the Fermi energy. Similarly, the density of states for holes $\bar\rho(\varepsilon)$ is given by $\delta(\varepsilon)$ for quantum dots and $\nu_i[1-f(\varepsilon)]$ for metallic systems. Finally, $P(\varepsilon)$ denotes the probability density that the tunneling electron exchanges the energy $\varepsilon$ with the environment.

For a simple tunnel junction without any embedded quantum dot system, there is no directed charge current in the absence of an applied bias voltage as the rectification of thermal fluctuations requires the presence of a nonlinearity in the heat engine. The simplest way to achieve such a nonlinearity is to add a single quantum dot or metallic island into the junction, cf. \fref{fig:EM_Harvesting}b). In the limit of strong Coulomb blockade, the quantum dot or metallic island is either empty or occupied with a single excess electron with energy $E$. The corresponding occupation probabilities $p_0$ and $p_1$ in the stationary state follow from a simple rate equation $0=(\Gamma_{\text{L}-}+\Gamma_{\text{R}+})p_1-(\Gamma_{\text{L}+}+\Gamma_{\text{R}-})p_0$ with $p_0+p_1=1$. Here, $\Gamma_{r\pm}$ denotes the transition rate for an electron tunneling left (right) through barrier $r$ evaluated according to \eref{eq:P(E)transition}.
In the following, we assume that the right tunnel barrier has a much larger transition rate than the left barrier. We, furthermore, assume that its capacitance is larger than that of the left junction such that it effectively decouples from the environment. Under these conditions, we obtain a directed charge current that is simply given by 
\begin{equation}
	I=f(V-E)\Gamma_{\text{L}+}-f(E-V)\Gamma_{\text{L}-},
\end{equation}
and thus finite even in the absence of an applied bias voltage. We remark that the above expression also describes the current through a junction between two quantum dots with energy levels $E_\text{L}$ and $E_\text{R}$ such that $E=E_\text{L}-E_\text{R}$ if the appropriate definitions for the electron and hole density of states are inserted in \eref{eq:P(E)transition} and the dot is preferably singly occupied.

By applying a bias voltage $V$ against the thermally driven current, we can generate a finite output power. As for any Coulomb-blockade heat engine, the output power is limited by the operation temperature. For a device operating at a temperature of about \unit[1]{K}, one obtains a power in the femtowatt range.

The efficiency of the heat engine depends strongly on the junction type. For a junction between two quantum dots with discrete energy levels, the tight-coupling limit is reached as every electron that is transferred between the dots has to absorb a photon with energy $E$. Consequently, the heat engine can reach Carnot efficiency and achieve a large efficiency at maximum power.
For junctions between a discrete level and a metal or between two metals, the situation is different. Now photons of different energies can be absorbed, leading to a significantly lower efficiency. Interestingly, the efficiency depends on whether the environment is hotter or colder than the electronic system. For a cold electron system, the sharp Fermi functions help to achieve unidirectional transport against the bias voltage such that decent efficiencies can be achieved. In contrast, for a hot electron system the thermally smeared Fermi function give rise to a current flow with the bias voltage such that only small efficiencies are possible.

\subsubsection{Microwave cavity photons}
\begin{figure}
	\centering\includegraphics[width=.475\columnwidth]{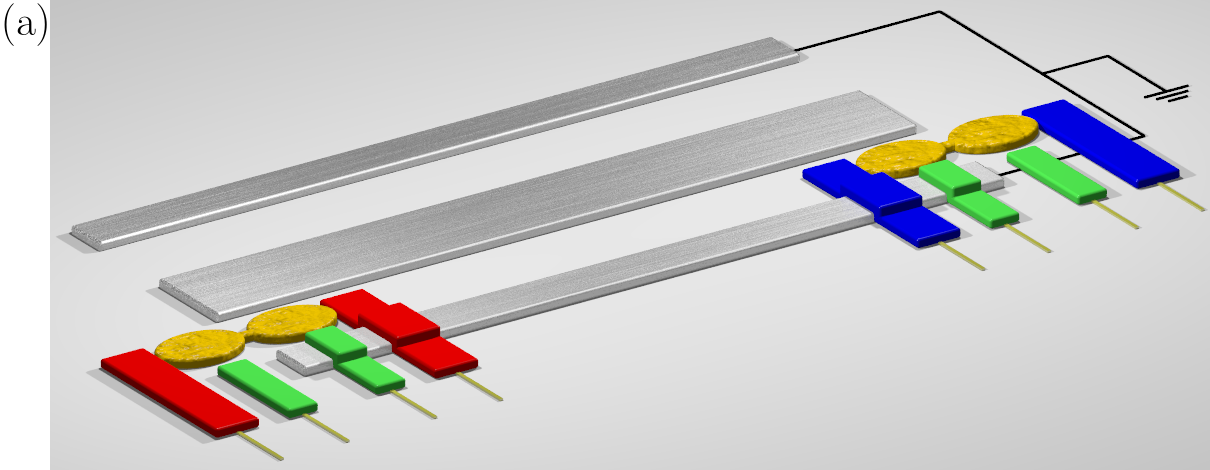}
	\includegraphics[width=.475\columnwidth]{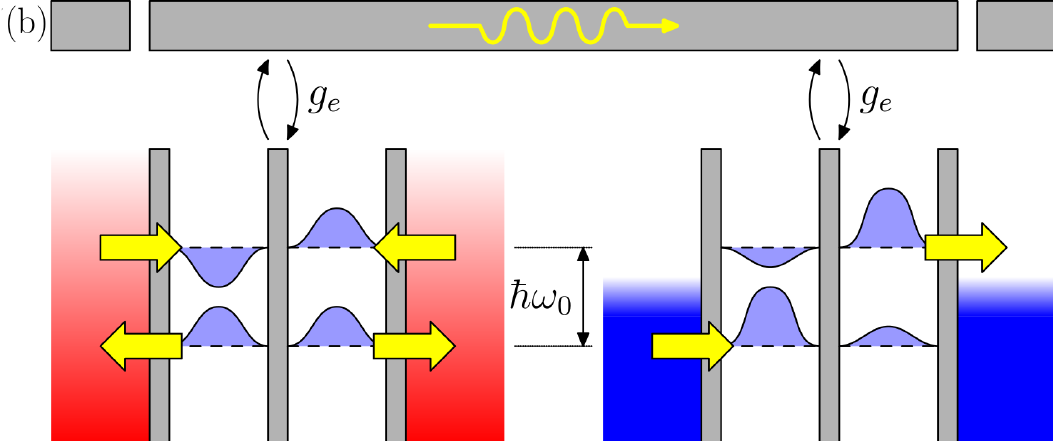}
	\caption{\label{fig:CQED_Model}(a) Sketch of the microwave cavity heat engine. Double quantum dots (yellow) are coupled to hot (red) or cold (blue) electronic reservoirs. The level positions of the dots are tunable via voltages applied to gates (green). The two double dots are connected via a superconducting microwave cavity (gray). (b) Operation scheme of the heat engine. Electrons enter the excited state of the hot DQD and leave the ground state after emitting a photon into the cavity. At the cold dot, electrons enter the ground state preferrably from the left and leave the excited state to the right after absorbing a photon from the cavity. As a result of this asymmetry, a directed charge current flows through the cold dot. (Reprinted with permission from~\cite{bergenfeldt_hybrid_2014}. Copyright 2014 American Physical Society.)}
\end{figure}

As a second example of a photon-driven heat engine we now consider a system where a superconducting microwave cavity connects two mesoscopic conductors, cf. \fref{fig:CQED_Model}. Such hybrid structures that allow the investigation the interplay of light and matter at the nansocale have recently gained a lot of interest both theoretically~\cite{childress_mesoscopic_2004,trif_spin_2008,jin_lasing_2011,cottet_subradiant_2012,cottet_microwave_2012,bergenfeldt_microwave_2012,bergenfeldt_nonlocal_2013,lambert_photon-mediated_2013,contreras-pulido_non-equilibrium_2013,xu_quantum_2013,xu_full_2013} as well as experimentally~\cite{frey_characterization_2011,delbecq_coupling_2011,frey_dipole_2012,frey_quantum_2012,petersson_circuit_2012,delbecq_photon-mediated_2013,toida_vacuum_2013,wallraff_comment_2013,basset_single-electron_2013} in the context of circuit quantum electrodynamics.
Similar to the previously discussed energy harvester, the hybrid microwave cavity heat engine~\cite{bergenfeldt_hybrid_2014} also allows to separate the hot and the cold part of the engine by a macroscopic distance of the order of a centimeter. This suppresses leakage heat currents far more efficiently than vacuum nanogaps that are only a few nanometer wide~\cite{taliashvili_vacuum_2013,jangidze_electroplating_2012}. In addition, the heat engine has the advantage that the cavity helps in efficiently transferring photons from the hot to the cold side whereas otherwise photons just get randomly emitted into all directions and can be lost for the energy harvesting purpose.

The heat engine consists of two double quantum dots $i=1,2$, connected to each other via a microwave cavity. Due to strong Coulomb interactions, the double dots can be either empty or occupied with a single electron. In the following, we neglect the electron spin as it only renormalizes the tunnel couplings in a trivial way. The eigenstates of the singly-occupied dot can be expressed as linear combinations of the left $\ket{L}_i$ and right $\ket{R}_i$ double quantum dot states as
\begin{eqnarray}
	\ket{+}_i=\cos\theta_i\ket{L}_i-\sin\theta_i\ket{R}_i,\\
	\ket{-}_i=\sin\theta_i\ket{L}_i+\cos\theta_i\ket{R}_i.
\end{eqnarray}
The mixing angle $\theta_i$ that characterizes the hybridization of the dot levels depends on the level positions and the interdot tunnel coupling. It can be controlled by gate voltages applied to the double quantum dot. The energy difference between the eigenstates $\ket{+}_i$ and $\ket{-}_i$ can be tuned independently of $\theta_i$ such that it matches the resonance frequency $\hbar\omega_0$ of the microwave cavity. The coupling between the double dots and the cavity with coupling strength $g_e$ then induces transitions between the ground and excited state of the dot accompanied by the absorption or emission of a microwave photon in the cavity.

Each quantum dot is furthermore tunnel coupled to two electronic reservoirs $\nu=L,R$ at temperature $T_i$ and chemical potential $\mu_{i\nu}$ with tunnel coupling strength $\Gamma_{i\nu}$. In the following, we assume for simplicity symmetric tunnel couplings for each double quantum dot, $\Gamma_{i\nu}=\Gamma_i$. We remark that the actual transition rates of the system depend not only on the tunnel coupling $\Gamma_i$ but also on the mixing angle $\theta_i$ that can effectively break the left-right symmetry within a double quantum dot. 

We describe transport through the system again within a generalized master equation approach that integrates out the noninteracting electronic reservoirs and characterizes the remaining quantum system consisting of the two double quantum dots and the microwave cavity by its reduced density matrix. In the limit of strong electron-photon coupling, $g_e\gg\Gamma_i$, coherences between different eigenstates of the quantum system can be neglected to lowest order in the tunnel coupling such that a simple rate equation description holds. It is this limit that we will consider first before taking into account the effects of finite electron-photon coupling as well as of relaxation and dephasing within the double quantum dots.

Before discussing the thermoelectric performance of the heat engine, we elucidate the basic operation principle that drives a charge current through the cold dot in the absence of any applied bias voltage. To this end, we choose a situation where the mixing angle of the hot dot is $\theta_1=\pi/4$. In this case, the eigenstates are the bonding and antibonding states. As they do not break left-right symmetry, there is no directed charge current through the double dot. For the cold double dot, we choose the mixing angle $\theta_2\neq\pi/4$ such that the ground state couples more strongly to one electrode while the excited states couples more strongly to the other one. Electrons then tunnel into the ground states of the cold double dot from one side, absorb a photon from the cavity to get into the excited state and then tunnel out to the other side. As there is a net flow of heat (and hence microwave photons) from the hot to the cold double dot, we thus obtain a finite directed charge current through the cold double dot. We remark that the direction of the charge current depends on the mixing angle as it determines the asymmetry of eigenstates. Unlike the energy dependence of tunnel couplings that is required for the heat engines discussed in Sec.~\ref{sec:CBfluct}, the asymmetry can therefore be changed in a controlled way by manipulating $\theta_2$ via gate voltages.

\begin{figure}
	\centering\includegraphics[width=\columnwidth]{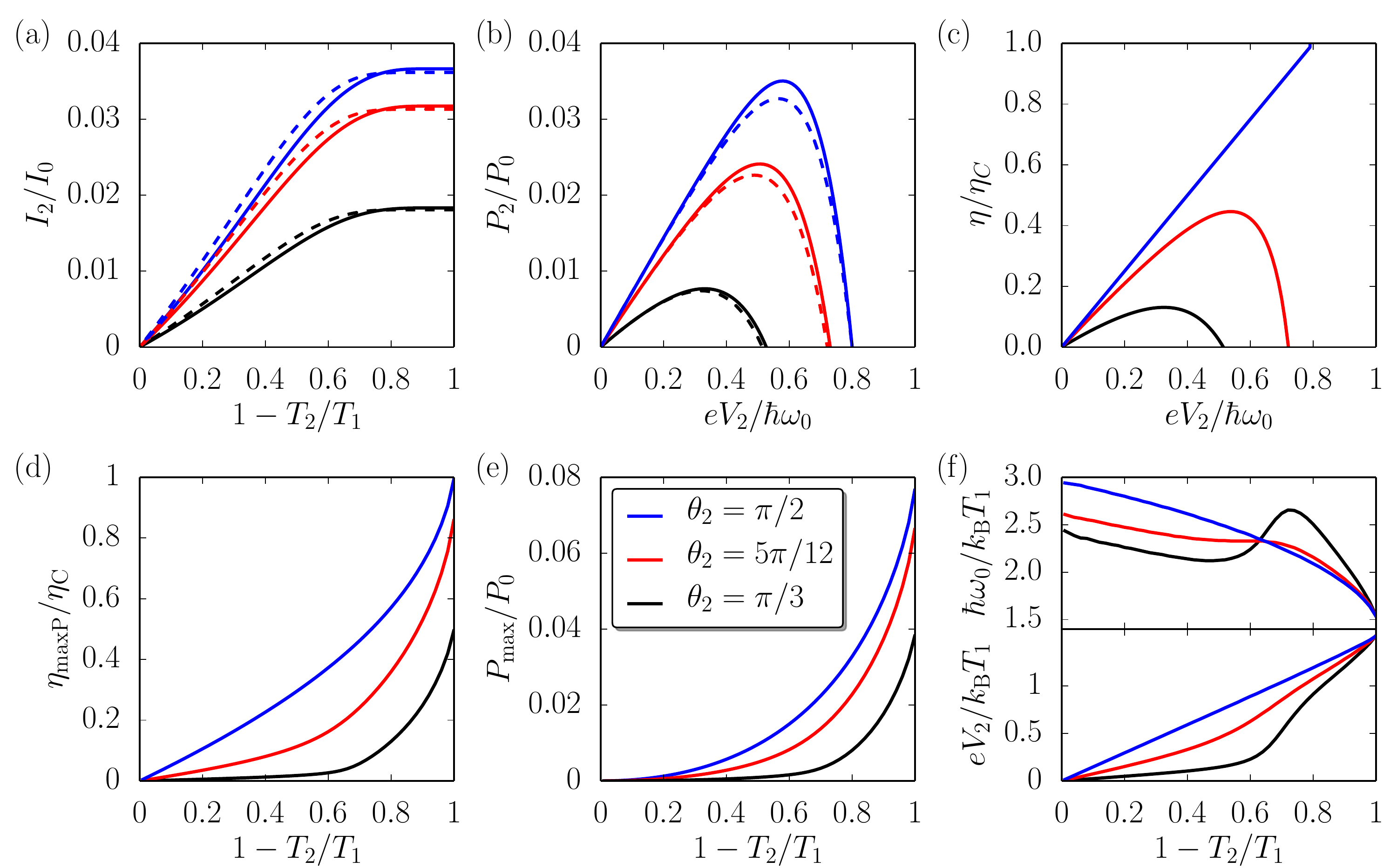}
	\caption{\label{fig:CQED_Results}Thermoelectric performance of the cavity heat engine. (a) Heat-driven charge current in units of $I_0=e\Gamma/2$ as a function of the temperature bias for different values of the mixing angle $\theta_2$. The analytic results~\eref{eq:CQED_current_analytic} (dashed lines) compare well with the full numerics (solid lines). Parameters are $\kB T_1=0.5\hbar\omega_0$ and $\Gamma=\Gamma_1=\Gamma_2$. (b) Output power in units of $P_0=\Gamma\kB T_1$ and (c) efficiency over Carnot efficiency as a function of the applied bias voltage. The analytical results for the power~\eref{eq:CQED_powerana} (dashed lines) agree well with the numerics (solid lines). Parameters are $\kB T_1=0.5\hbar \omega_0$ and $\kB T_2=0.1\hbar\omega_0$. (d) Efficiency at maximum power and (e) maximum power as a function of the temperature bias. The optimal values of the frequency $\omega_0$ and voltage $V_2$ are shown in (f). (Reprinted with permission from~\cite{bergenfeldt_hybrid_2014}. Copyright 2014 American Physical Society.)}
\end{figure}

The thermoelectric performance of the microwave cavity heat engine is summarized in \fref{fig:CQED_Results}. In panel a) we show that charge current $I_2$ that flows through double quantum dot 2 without an applied bias voltage. It grows linearly with the applied temperature bias and saturates for $T_2\ll T_1$. For $\theta_2\to\pi/2$, the tight-coupling limit is reached where one electron is transferred through the cold double dot for each photon that is transferred through the cavity. As a consequence, we find that the charge current $I_2$ is proportional to the heat current $J_1$ flowing out of the hot reservoirs, $I_2/J_1=e/(\hbar\omega_0)$, where the ratio of these two currents is determined only by the electron charge and the photon energy. In the regime where the cavity is mostly empty, i.e., $\bar f_i=f_{Li}(\hbar\omega_0/2)=f_{Ri}(\hbar\omega/2)\ll1$ the charge current is approximately given by the analytic expression 
\begin{equation}\label{eq:CQED_current_analytic}
I_{2}=\cos(2\theta_{2})\frac{e\Gamma_{1}\Gamma_{2}}{\Gamma_{1}+\Gamma_{2}}(\bar{f}_{2}^{2}-\bar{f}_{1}^{2}),
\end{equation}
that agrees well with the full numerical result, cf. the dashed lines in \fref{fig:CQED_Results}a). As can be inferred from \eref{eq:CQED_current_analytic}, the current becomes largest when $\Gamma_1=\Gamma_2\equiv \Gamma$. We will therefore focus on this limit in the following discussion.

The output power against an externally applied bias voltage $V_2$ is depicted in \fref{fig:CQED_Results}b). It shows a typical behaviour by growing from $P=0$ at $V_2=0$ to a maximal value and dropping to zero at the stopping voltage. For $\bar f_{\nu i}=f_{\nu i}(\hbar \omega_0/2)\ll1$ it is well approximated by
\begin{equation}\label{eq:CQED_powerana}
\eqalign{
P_2=\frac{e\Gamma V_2}{2}\left[\sin(2\theta_{2})\{\bar{f}_{L2}-\bar{f}_{R2}\}-\cos(2\theta_{2})\bar{f}_{1}^{2}\vphantom{\frac{1}{2}}\right.\cr\left.
+\cos(2\theta_{2})\bar{f}_{L2}\bar{f}_{R2}+\frac{\sin^{2}(2\theta_{2})}{4}\{\bar{f}_{L2}^{2}-\bar{f}_{R2}^{2}\}\right].}
\end{equation}
The efficiency in general shows behaviour qualitatively similar to the output power. Only in the tight-coupling limit it instead grows linearly with the applied bias voltage and reaches Carnot efficiency at the stopping voltage. The results for the maximum power $P_\text{max}$ and the associated efficiency at maximum power $\eta_\text{maxP}$ together with the optimizing values of $\omega_0$ and $V_2$ are shown in \fref{fig:CQED_Results}d)-f). In the tight-coupling limit, $\eta_\text{maxP}$ grows as $\eta_C/2$ in the linear response regime. In the nonlinear regime, it grows more quickly and reaches $\eta_C$ for $T_1\gg T_2$. For $\theta_2<\pi/2$, the efficiency at maximum power is slightly reduced but exhibits a qualitatively similar behaviour.

\begin{figure}
	\centering\includegraphics[width=.5\columnwidth]{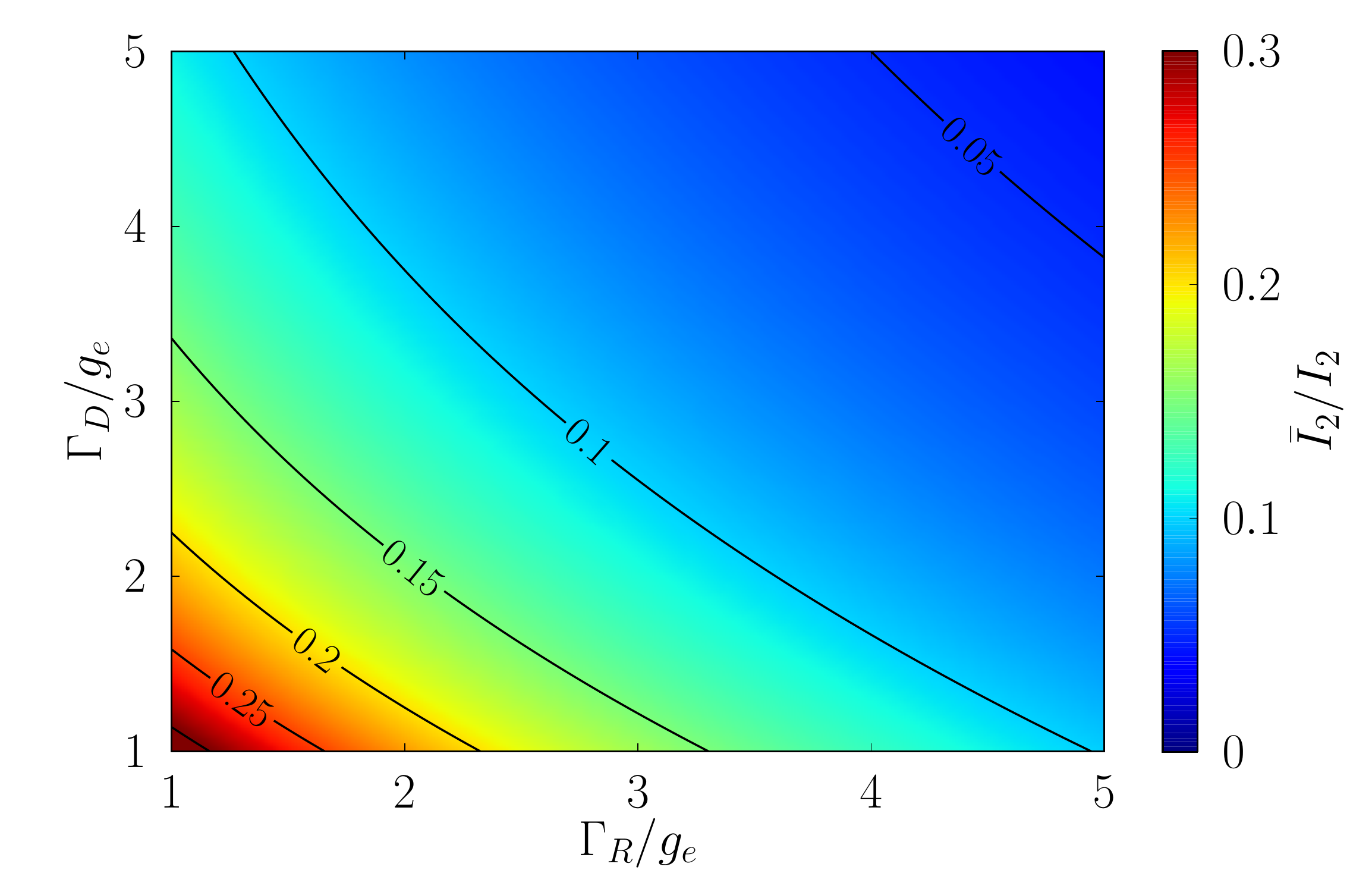}
	\caption{\label{fig:CQED_Dephasing}Influence of electronic relaxation and dephasing rates on the heat-driven current. In the optimal case, the current takes the value $I_2=e\Gamma/2$. (Reprinted with permission from~\cite{bergenfeldt_hybrid_2014}. Copyright 2014 American Physical Society.)}
\end{figure}

In current experiments on circuit quantum electrodynamics relaxation and dephasing within the double quantum dots are the major obstacles in reaching the strong coupling limit. When taking into account a finite value of the electron-photon coupling $g_e$ together with relaxation and dephasing processes with rates $\Gamma_R$ and $\Gamma_D$, respectively, we find that the charge current through the system is reduced to
\begin{equation}\label{eq:reCurr}
\bar{I}_{2}=\frac{4 g_{e}^{2}}{(\Gamma+\Gamma_{R})(\Gamma+\Gamma_{R}+4\Gamma_{D})+4g_{e}^{2}}I_{2}.
\end{equation}
The corresponding current suppression is shown in \fref{fig:CQED_Dephasing} as a function of $\Gamma_R$ and $\Gamma_D$. Even when taking these imperfections into account, we estimate that for realistic system parameters one can achieve charge currents of the order of $\unit[0.2]{pA}$ clearly within the reach of current experiments.

\section{\label{sec:concl}Conclusions}
In this review we discussed different types of thermoelectric energy harvesters based on multi-terminal quantum dot and well setups. We presented nano heat engines based on Coulomb-coupled conductors. In particular, we looked at a setup based on Coulomb-blockade quantum dots that turned out to be ideally efficient but only gives small currents and powers. A similar system based on chaotic cavities was shown to yield much larger currents but had a significantly reduced efficiency. The optimal heat engine that has both a large output power in addition to a good efficiency was then found to be based on resonant tunneling through quantum dots or quantum wells. In the second part of this review, we discussed different types of heat engines that are powered by absorbing bosons from the environment.
We analyzed heat engines based on molecular junctions where electrons couple to phononic degrees of freedom from which energy is harvested. We then turned to a related setup that is driven by magnons from a ferromagnetic insulator. This type of setup allows one to drive pure spin currents as well as spin-polarized charge currents and therefore makes connection to the emerging field of spin caloritronics. Finally, we discussed setups that harvest microwave photons from the electromagnetic environment or from a superconducting microwave cavity thus providing a bridge between energy harvesting and circuit quantum electrodynamics.

The field of energy harvesting with mesoscopic conductors of course still faces a number of open questions.
While most of the setups presented in this review were described in terms of minimal models, it would be interesting to have more realistic descriptions that include, e.g., charging effects in quantum wells or provide a microscopic model of heat injection into the central region of the resonant tunneling heat engines.
Furthermore, the influence of phonons has to be better understood as they not only degrade the efficiency but also make it hard to maintain a given temperature bias across the device. In particular, it would be desirable to investigate systems where an analytic treatment of their influence can be made and to invent setups where they can be controlled in a systematic way.
Another interesting question addresses the thermoelectric performance of multi-terminal heat engines with broken time-reversal symmetry. Theoretically, these devices allow to get arbitrarily close to Carnot efficiency at finite output power. But how close can any physical realization actually get?
Finally, the field of mesoscopic energy harvesting will greatly benefit from experiments that demonstrate that the theoretical ideas presented in this review can  also be put into practice.

%Acknowledgments
%
\ack
Above all, we are deeply indebted to the support and inspiration of the late Markus Büttiker during many years of collaborations that led to our works presented in this review.
We furthermore acknowledge numerous fruitful discussions with
C. Bergenfeldt, 
H. Buhmann, 
C. Flindt, 
F. Hartmann, 
P. Jacquod, 
A. M. Lunde,
R. López, 
I. Martin,
L. Molenkamp, 
M. Moskalets, 
P. Samuelsson, 
D. Sánchez, 
J. Splettstoesser, 
H. Thierschmann, 
R. Whitney 
and
L. Worschech.

We also acknowledge financial support from the European STREP project NANOPOWER, the Swiss NSF via the NCCR QSIT, the Spanish MICINN Juan de la Cierva program and MAT2011-24331, the COST Action MP1209 and the NSF grant DMR-0844899.

\section*{References}
% \bibliographystyle{iopart-num}
% \bibliography{/home/bjoern/LaTeX/Bibtex/Meine_Bibliothek}

\providecommand{\newblock}{}

\end{document}